\documentclass[11pt]{article}
\usepackage{fullpage,epsf,amssymb,amsthm,amsfonts,amsmath,latexsym,array,graphicx,extarrows,mathtools,mathstyle,mathrsfs,slashed,subcaption,amsbsy,csquotes,accents}
\usepackage[normalem]{ulem}
\usepackage{authblk}
\usepackage{cite}
\usepackage{color}    
\usepackage{float} 
\usepackage{hyperref} 
\usepackage[export]{adjustbox}
\usepackage{hhline}

\numberwithin{equation}{section} 

\renewcommand{\tfrac}[2]{\genfrac{}{}{}{\textstyle}{#1}{#2}}

\title{\bf{Resonances at finite temperature from the lattice}} 

\author[]{Jakob Hoffmann}
\author[]{Peter Lowdon}
\author[]{Owe Philipsen}

\affil[]{{\scriptsize Institut f\"{u}r Theoretische Physik, Goethe-Universit\"{a}t, Max-von-Laue-Str. 1,  60438 Frankfurt am Main, Germany}}

\date{}
\begin{document}
\maketitle

\begin{abstract}
\noindent
The properties of hadronic resonances at finite temperature constitute an important probe of the thermal QCD medium. In this work we use the concept of thermoparticles, which characterise thermally-modified but stable particle-like states, to define the notion of two-particle scattering at finite temperature, and establish a unitarity relation for the thermal scattering amplitude. We derive a finite-temperature generalisation of the vacuum two-particle quantisation condition via a skeleton expansion of the finite-volume correlation function. Solving this condition at the finite-volume energy levels of the system constrains the form of the thermal scattering amplitude, and hence the properties of resonances. In contrast to the vacuum case, the kinematic function containing the leading finite-volume corrections is finite at all energies, reflecting the fact that thermoparticles have broadened spectral peaks due to their interactions with the thermal medium. Since all lattice simulations involve a finite temporal extent, our approach can also be used to study finite-temporal size effects in vacuum analyses.
\end{abstract}

\newpage

\section{Introduction}
\label{sec1}

Resonances are expected to experience major modifications in a medium of strongly-interacting matter at finite temperature. In the case of vector mesons, these effects can be probed experimentally by the emitted dilepton spectrum~\cite{Kapusta:2006pm}. Theoretical studies of these non-perturbative thermal states are so far limited to low-energy effective models~\cite{Urban:1999im,Jung:2016yxl}, chiral perturbation theory extended to low temperatures~\cite{Dobado:2002xf,GomezNicola:2023rqi}, and functional approaches applied to QCD directly~\cite{Gao:2020hwo}. In this work, we propose a finite-temperature generalisation of the existing finite-volume formalism used to extract resonance properties in vacuum from lattice QCD. The vacuum formalism has been successfully applied to study a wide variety of hadronic resonances from first principles~\cite{Briceno:2017max,Hansen:2019nir,Mai:2022eur}. This is particularly challenging since resonances decay into multi-particle states, which means their masses and decay widths are not associated to one specific energy level, as is the case for stable bound states. Instead, hadronic resonances are identified by poles on the second Riemann sheet of the real-time infinite-volume scattering amplitude, and are hence not directly accessible in Euclidean lattice simulations. State-of-the-art spectroscopic studies in lattice QCD use finite-volume formulations of multi-particle scattering states in order to extract information about resonances. Such formulations contain quantisation conditions that relate the finite-volume energy levels computed on the lattice to the infinite-volume multi-particle scattering amplitude, which is therefore constrained by numerical data. Multiple versions of the quantisation condition have been derived using different approaches~\cite{Luscher:1986Scat,Luscher:1991ux,Rummukainen:1995vs,Kim:2005gf,He:2005ey,Fu:2012tj,
Leskovec:2012xz,Gockeler:2012dk,Hansen:2012tf,Briceno2014,Hansen:2014eka,Hansen:2015zga,Mai:2017vot,Mai:2017bge,Doring:2018xxx,Blanton:2019igq,Hansen:2020zhy,Blanton:2021mih,Draper:2023xvu,Draper:2024qeh,Dawid:2024oey,Schaaf:2025pnf,Hansen:2025oag,Dawid:2025wsn,Dawid:2025zxc,Briceno:2025yuq}, and applied to extract the properties of a range of resonances, including those decaying into two hadrons~\cite{Dudek:2012xn,Wilson:2014cna,Wilson:2015dqa,Briceno:2017qmb,Wilson:2019wfr,Gayer:2021xzv,Boyle:2024grr,Lang:2025pjq,Hu:2017wli,Andersen:2017una,Bulava:2022vpq,
BaryonScatteringBaSc:2023ori,BaryonScatteringBaSc:2023zvt,Leskovec:2018lxb,Alexandrou:2023cqg}, and in some cases three hadrons~\cite{Mai:2021nul,Yan:2024gwp,Dawid:2025doq}. The key idea in the derivation of the quantisation conditions is to identify the leading-order finite-volume corrections, as argued in Refs.~\cite{Kim:2005gf,Hansen:2012tf}. In vacuum, these leading-order contributions are generated exclusively by combinations of poles from the single-particle on-shell propagators, whilst the non-singular components lead to exponentially-suppressed finite-volume corrections. \\

\noindent
The main challenge at finite temperature is the presence of a thermal medium, which has a significant influence on the states of the theory. At zero temperature, the asymptotic scattering states correspond to free on-shell particles, whose propagator poles dominate the finite-volume corrections within the quantisation condition. However, at finite temperature the thermal medium is everywhere-present, even at asymptotically-large times, and hence scattering states cannot be defined by free on-shell vacuum particles~\cite{Landsman:1988ta}. In Refs.~\cite{Bros:1992ey,Buchholz:1993kp} the concept of thermoparticles was introduced in order to resolve these inconsistencies. Thermoparticles represent thermally-damped but stable particle-like excitations, which reduce to on-shell stable particles in the zero-temperature limit. These excitations dominate the correlation functions of the theory at large times~\cite{Bros:2001zs}, and are therefore natural candidates for describing scattering states at finite temperature. In this work we use thermoparticles in order to define the basic concepts of scattering at finite temperature, and to generalise the finite-volume vacuum quantisation conditions by following an analogous approach to Refs.~\cite{Kim:2005gf,Hansen:2012tf}. An important feature of the resulting finite-volume corrections is that the singularities in the vacuum case get smoothly regularised by the inclusion of finite-temporal size $L_{\tau}$ effects, and this allows for a controlled approach of the vacuum ($L_{\tau}\rightarrow \infty$) limit. \\

\noindent
The paper is structured as follows. In Secs.~\ref{sec:TP_basics} and~\ref{sec:infinite-volume-scattering} we outline the characteristics of thermoparticles and thermal scattering amplitudes, in Sec.~\ref{sec:qc2-derivation} we derive a finite-temperature generalisation of the two-particle quantisation condition, and present numerical results, and in Sec.~\ref{sec:Conclusion} we summarise our main findings and discuss potential future applications and extensions.

\section{Thermoparticles and their characteristics}
\label{sec:TP_basics}

In order to establish the connection between scattering amplitudes and the properties of resonances at non-vanishing temperatures $T=1/\beta>0$ one needs to understand the fundamental constraints satisfied by thermal correlation functions. For infinite-volume QFTs the notion of thermal equilibrium is defined by the Kubo-Martin-Schwinger (KMS) condition, where for scalar fields the correlation functions satisfy~\cite{Haag:1967sg}  
\begin{align}
&\langle \Omega_{\beta}|\phi(x_{1})\cdots \phi(x_{k})\phi(x_{k+1})\cdots \phi(x_{n})|\Omega_{\beta}\rangle  \nonumber \\
& \quad\quad\quad\quad\quad\quad\quad\quad\quad\quad = \langle \Omega_{\beta}|\phi(x_{k+1})\cdots \phi(x_{n}) \phi(x_{1}+i(\beta,\vec{0}))\cdots \phi(x_{k}+i(\beta,\vec{0}) )|\Omega_{\beta}\rangle.
\label{KMS}
\end{align}
In contrast to the vacuum, the thermal equilibrium state $|\Omega_{\beta}\rangle$ no longer represents a unique ground state of the system, and this has important implications for the spectrum of the theory. In particular, it follows that the thermal spectrum must be symmetric around zero, and hence for any state $|\omega\rangle$ with positive energy $\omega>0$ there exists a corresponding \textit{negative} energy state $|-\omega\rangle$~\cite{Landsman:1988ta}. At $T=0$ such states are excluded by the relativistic spectral condition~\cite{Streater:1989vi,Haag:1992hx,Bogolyubov:1990kw}, which asserts that $\omega \geq 0$ and $p^{2} \geq 0$. However, when $T>0$ the appearance of negative-energy states represents a physical effect, namely the extraction of energy from the thermal medium, which is equivalent to the creation of hole-like states~\cite{Bros:2001zs}. Although these states exist, the probability for their creation is thermodynamically suppressed, since the KMS condition in Eq.~\eqref{KMS} relates the momentum-space two-point function $\widetilde{\mathcal{W}}(\omega,\vec{p})$ and spectral function $\rho(\omega,\vec{p})$ via          
\begin{align}
\widetilde{\mathcal{W}}(\omega,\vec{p}) = \frac{\rho(\omega,\vec{p})}{1-e^{-\beta \omega}},
\label{W_rep} 
\end{align}
which implies: $\widetilde{\mathcal{W}}(\omega,\vec{p}) \sim e^{-\beta |\omega|}$ for $\omega \rightarrow -\infty$. In the zero-temperature limit $\beta \rightarrow \infty$ Eq.~\eqref{W_rep} reduces to $\widetilde{\mathcal{W}}_{\text{vac}}(\omega,\vec{p}) = \theta(\omega) \, \rho(\omega,\vec{p})$, which reflects the positive-energy constraint imposed by the relativistic spectral condition in the vacuum theory, as expected. \\

\noindent
Another fundamental constraint on thermal correlation functions is causality, which requires that the field commutator satisfies the condition: $\left[\phi(x),\phi(y)\right]=0$ for $(x-y)^{2}<0$. Since the spectral function $\rho(\omega,\vec{p})$ is defined as the Fourier transform of $\langle \Omega_{\beta}|\left[\phi(x),\phi(y)\right]|\Omega_{\beta}\rangle$, in Ref.~\cite{Bros:1996mw} the authors proved that the causality condition implies that $\rho(\omega,\vec{p})$ satisfies the general representation\footnote{ For a more recent discussion of this representation see Ref.~\cite{Nair:2025jgl}.} 
\begin{align}
\rho(\omega,\vec{p}) = \int_{0}^{\infty} \! ds \int \! \frac{d^{3}\vec{u}}{(2\pi)^{2}} \ \epsilon(\omega) \, \delta\!\left(\omega^{2} - (\vec{p}-\vec{u})^{2} - s \right)\widetilde{D}_{\beta}(\vec{u},s).    
\label{rho_rep}
\end{align}
From this representation it follows that the fundamental excitations of the medium are encoded in the structure of the thermal spectral density $\widetilde{D}_{\beta}(\vec{u},s)$. Given that the vacuum theory contains a stable particle state of mass $m$, there are general theoretical arguments~\cite{Bros:1996mw,Bros:1992ey,Bros:2001zs}, as well as concrete numerical evidence\footnote{This includes real~\cite{Lowdon:2024atn,Ali:2026ehk} and complex~\cite{Lowdon:2025fyb,Lowdon:2025ait} scalar theories, as well as QCD~\cite{Lowdon:2022xcl,Bala:2023iqu}.}, that $\widetilde{D}_{\beta}(\vec{u},s)$ contains a distinguished component of the form 
\begin{align}
\widetilde{D}_{m,\beta}(\vec{u})\, \delta(s-m^{2}).
\label{TP_damping}
\end{align}
In the zero-temperature limit: $\widetilde{D}_{m,\beta}(\vec{u}) \rightarrow (2\pi)^{3}\delta^{3}(\vec{u})$, and hence the spectral function of a stable particle state in vacuum $\rho_{\text{vac}}(\omega,\vec{p})= 2\pi \epsilon(\omega)\delta(p^{2}-m^{2})$ is recovered. Equation~\eqref{TP_damping} therefore represents the finite-temperature generalisation of a stable particle state. In order to draw a clear distinction with other potential thermal excitations, these components were subsequently referred to as \textit{thermoparticles}~\cite{Buchholz:1993kp}. \\

\noindent
Due to the representation in Eq.~\eqref{rho_rep}, and the specific form of Eq.~\eqref{TP_damping}, thermoparticles have a number of distinctive properties. These properties are discussed extensively in Ref.~\cite{Ali:2026ehk}, but we will briefly summarise the main ones below: 

\begin{itemize}

\item  The non-trivial $\vec{u}$-dependence of $\widetilde{D}_{m,\beta}(\vec{u})$ implies that the thermoparticle spectral function peak is broadened around the vacuum $p^{2}=m^{2}$ singularity, which captures the effects of collisional interactions with the thermal medium. In position space $D_{m,\beta}(\vec{x})$ reduces the propagation amplitude of the state with increasing temperature, lowering its mean-free path, and therefore has the interpretation of a thermal damping factor.

\item Thermoparticle spectral functions $\rho_{\text{TP}}(\omega,\vec{p})$ have an energy threshold at $|\omega|=m$, and hence one can write
\begin{align}
\rho_{\text{TP}}(\omega,\vec{p}) = \theta(\omega^{2}-m^{2})\varrho_{\text{TP}}(\omega,\vec{p}).
\label{TP_thresh}
\end{align}
The thermal medium must therefore be excited with an energy $\omega \geq m$ in order to create a thermoparticle state with non-vanishing probability.

\item Thermoparticles dominate the large-time $|x_{0}| \rightarrow \infty$ behaviour of thermal correlation functions, and therefore provide a consistent description of finite-temperature scattering states. These states are no longer on shell, like in the vacuum case, which reflects the fact that even at large times the thermal medium contains interactions~\cite{Ali:2026ehk}. The corresponding damping factor of these states $D_{m,\beta}(\vec{x})$ is \textit{uniquely} fixed by the dynamics~\cite{Bros:2001zs,Bros:2003zs}, and hence the properties of these states differ between specific theories. This is a manifestation of the property that finite-temperature phenomena are strongly dependent on the dynamics of the underlying medium.

\end{itemize} 

\noindent
Given the various characteristics of thermoparticles it is natural that these degrees of freedom form the basis of scattering amplitudes at finite temperature. In particular, if the temperature of the system is not too large one would expect thermoparticles to dominate the low-energy spectrum of the theory\footnote{Strong evidence of thermoparticle dominance has already been found in the light pseudo-scalar meson spectrum in QCD~\cite{Lowdon:2022xcl,Bala:2023iqu} for temperatures around the pseudo-critical temperature $T_{\text{pc}}$.}, and hence any skeleton-like expansion should be parametrised in terms of thermoparticle propagators. In general, the analytic thermoparticle propagator $\widetilde{G}_{\text{TP}}(k_{0},\vec{p})$ can be written in the form
\begin{align}
\widetilde{G}_{\text{TP}}(k_{0},\vec{p}) = -\int_{-\infty}^{\infty} \frac{dq_{0}}{2\pi}\frac{\rho_{\text{TP}}(q_{0},\vec{p})}{k_{0}-q_{0}},
\label{G_TP}
\end{align}
where taking $k_{0} \rightarrow \omega \pm i\epsilon$ recovers the real-time retarded and advanced propagators, and the imaginary-time propagator is obtained by setting $k_{0}=i\omega_{N}$, where $\omega_{N}=\tfrac{2\pi N}{\beta}$ ($N \in \mathbb{Z}$) are the discrete Matsubara frequencies~\cite{Kapusta:2006pm,Bellac:2011kqa}. The representation in Eq.~\eqref{G_TP} will be used frequently throughout this work\footnote{Technically, the existence of a \textit{single} analytic propagator $\widetilde{G}(k_{0},\vec{p})$ whose limits $k_{0} \rightarrow \omega \pm i\epsilon$ recover the retarded and advanced propagators requires that the spectral function must vanish in some non-vanishing energy-momentum region~\cite{Bros:1996mw}.}.

\section{Scattering at finite temperature} 
\label{sec:infinite-volume-scattering}

In this study we focus on the elastic scattering of two identical scalar states at finite temperature. At $T=0$ the incoming and outgoing particles have four-momenta $p_{1},p_{2}$ and $p'_{1},p'_{2}$, respectively, with energies $\omega_{i}= \sqrt{|\vec{p}_{i}|^{2}+m^{2}}$, $\omega'_{i}= \sqrt{|\vec{p}_{i}\hspace{0.1mm}'|^{2}+m^{2}}$ ($i=1,2$), and total energy $E=\omega_{1}+\omega_{2}$ and momentum $\vec{P}=\vec{p}_{1}+\vec{p}_{2}$. The two-body system can be described by the Lorentz-invariant Mandelstam variables $s=(p_1+p_2)^2$ and $t=(p_1-p'_1)^2$. When $T>0$ the boost invariance of the system is lost, which implies that scattering processes are no longer entirely fixed by $s$ and $t$, but depend on the frame via $P=(E,\vec{P})$. Nevertheless, since both $E$ and $\vec{P}$ remain conserved for elastic scattering, the process can be entirely parametrised in terms of the variables $(p_1,p'_1,E,\vec{P})$. As outlined in Sec.~\ref{sec:TP_basics}, thermoparticles are a natural candidate for describing scattering states at finite temperature. In this section we will explore the impact that these states have on the structure of thermal scattering amplitudes. 

\subsection{Thermal S-matrix and scattering states} 
\label{subsec:S-matrix}
 
A fundamental constraint on scattering at finite-temperature arises from the Narnhofer-Thirring theorem (NRT) theorem~\cite{Narnhofer:1983hp}, which implies that asymptotic thermal states with purely real dispersion relations result in a trivial thermal S-matrix, i.e. $S=1$. Consequently, on-shell vacuum states are not good candidates for describing scattering states at finite temperature. By contrast, thermoparticles have spectral functions with peaks which are broadened around the vacuum singularity $p^{2}=m^{2}$, and hence these states are manifestly off shell and avoid the constraints from the NRT theorem. Similarly to the vacuum case, one can also in principle construct multi-thermoparticle states as a product of single-thermoparticle states, e.g. $\left|p_1,p_2\right>_{\text{TP}}=\left|p_1\right>_{\text{TP}}\otimes \left|p_2\right>_{\text{TP}}$. If these states span the full Hilbert space $\mathcal{H}_{\beta}$ of the theory, as in the vacuum case~\cite{Bogolyubov:1990kw}, it follows in an analogous manner that the thermal S-matrix will be unitary, and hence $SS^{\dagger} =1$.   \\

\noindent
In vacuum theories one implicitly assumes the scattering states are on shell and have positive energy. This can be guaranteed by defining the one-particle states of mass $m$ as\footnote{See Ref.~\cite{Lorce:2019sbq} for a more in-depth discussion of these states.}    
\begin{align}
|p; m \rangle = 2\pi \,\theta(\omega)\,\delta(p^{2}-m^{2}) |p\rangle,
\label{p_on-shell}
\end{align}   
where $|p\rangle$ has unrestricted four-momentum $p$. Since the covariant pre-factor in Eq.~\eqref{p_on-shell} is the two-point function $\widetilde{\mathcal{W}}_{\text{vac}}(\omega,\vec{p})$ of a massive vacuum particle state, due to the KMS constraint in  Eq.~\eqref{W_rep} this suggests that for thermoparticles these states should be defined as 
\begin{align}
|p \rangle_{\text{TP}} = \frac{\rho_{\text{TP}}(\omega,\vec{p})}{1-e^{-\beta \omega}} |p\rangle.
\label{p_TP}
\end{align} 
In the zero-temperature limit ($\beta \rightarrow \infty$) it follows that
\begin{align}
|p \rangle_{\text{TP}} \rightarrow \theta(\omega)\rho_{\text{vac}}(\omega,\vec{p}) = |p; m \rangle,
\end{align}
and hence the one-particle vacuum scattering state is recovered, as expected.

\subsection{Thermal Optical Theorem} 

Using the definition of the asymptotic thermoparticle states in Eq.~\eqref{p_TP} one can explore the consequences of the unitarity of the thermal S-matrix, in particular how the optical theorem for vacuum QFTs~\footnote{See Ref.~\cite{Peskin:1995ev} for a detailed discussion of the optical theorem in the standard vacuum case.} generalises to finite temperature. Analogously to the vacuum case one can write $S=1+iT$, where the thermal $T$-matrix encodes all of the non-trivial scattering information. Since $SS^{\dagger}=1$, this implies the condition 
\begin{align}
T-T^{\dagger}=i\, TT^{\dagger}.
\label{eq:T-matrix-unitarity}
\end{align}
By defining the two-thermoparticle scattering amplitude $\mathcal{M}_{\text{TP}}(p_1,p_2;p'_1,p'_2)$ via
\begin{equation}
{}_{\text{TP}} \langle p'_1,p'_2|T|p_1,p_2 \rangle_{\text{TP}} =(2\pi)^4\, \delta^{(4)}(p'_1+p'_2-p_1-p_2)\,\mathcal{M}_{\text{TP}}(p_1,p_2;p'_1,p'_2),
\label{eq:def-scattering-amplitude}
\end{equation}
and inserting both operator representations in Eq.~\eqref{eq:T-matrix-unitarity} into the matrix element of two-thermoparticle states, one obtains a non-trivial relation between these scattering amplitudes. In particular, inserting the left-hand-side operator one obtains
\begin{align}
{}_{\text{TP}} \langle p'_1,p'_2|T-T^{\dagger}|p_1,p_2 \rangle_{\text{TP}} &= (2\pi)^4 \, \delta^{(4)}(p'_1+p'_2-p_1-p_2) \nonumber \\
& \quad\quad\quad \times \left(\mathcal{M}_{\text{TP}}(p_1,p_2;p'_1,p'_2)-\mathcal{M}_{\text{TP}}^{*}(p'_1,p'_2;p_1,p_2)\right) \!,
\label{eq:lhs-unitarity}
\end{align}
where $\mathcal{M}^{*}$ denotes the complex conjugate of $\mathcal{M}$, and from the right-hand-side operator 
\begin{align}
{}_{\text{TP}} \langle p'_1,p'_2|TT^{\dagger}|p_1,p_2 \rangle_{\text{TP}} &= \sum_{n}\left(\frac{1}{n!}\prod_{i=1}^{n}\int \frac{d^{4} q_{i}}{(2\pi)^{4}} \frac{\rho_{\text{TP}}(\omega_{i},\vec{q}_{i})}{1-e^{-\beta \omega_{i}}} \right) {}_{\text{TP}} \langle p'_1,p'_2|T|\{q_{i,(n)}\} \rangle_{\text{TP}}  \ {}_{\text{TP}} \langle \{q_{i,(n)}\}|T^{\dagger}|p_1,p_2 \rangle_{\text{TP}}, \nonumber \\
&= \sum_{n}\left(\frac{1}{n!}\prod_{i=1}^{n}\int \frac{d^{4} q_{i}}{(2\pi)^{4}} \frac{\rho_{\text{TP}}(\omega_{i},\vec{q}_{i})}{1-e^{-\beta \omega_{i}}} \right) (2\pi)^{8}\, \delta^{(4)}({\scriptstyle{\sum_{i=1}^{n}}}q_{i}-p_1-p_2)  \nonumber \\
&  \quad \times  \delta^{(4)}(p'_1+p'_2-{\scriptstyle{\sum_{i=1}^{n}}} q_{i}) \mathcal{M}_{\text{TP}}(\{q_{i,(n)}\};p'_1,p'_2) \mathcal{M}_{\text{TP}}^{*}(\{q_{i,(n)}\};p_1,p_2).
\label{eq:rhs-unitarity}
\end{align}
In the first line we inserted a complete set of $n$-thermoparticle states, which are assumed to span $\mathcal{H}_{\beta}$. We define $|\{q_{i,(n)}\} \rangle_{\text{TP}}= |q_{1} \rangle_{\text{TP}}\otimes \cdots \otimes |q_{n} \rangle_{\text{TP}}$ and include an explicit $1/n!$ symmetrisation factor due to the bosonic nature of these states. The unitarity relation in Eq.~\eqref{eq:T-matrix-unitarity} therefore implies the equality
\begin{align}
\mathcal{M}_{\text{TP}}(p_1,p_2;p'_1,p'_2)-\mathcal{M}_{\text{TP}}^{*}(p'_1,p'_2;p_1,p_2) &=  \sum_{n}\left(\frac{1}{n!}\prod_{i=1}^{n}\int \frac{d^{4} q_{i}}{(2\pi)^{4}} \frac{\rho_{\text{TP}}(\omega_{i},\vec{q}_{i})}{1-e^{-\beta \omega_{i}}} \right)    \nonumber \\
& \quad\quad \times  i(2\pi)^{4}\, \delta^{(4)}(p'_1+p'_2-{\scriptstyle{\sum_{i=1}^{n}}} q_{i}) \nonumber \\
&  \quad\quad \times \mathcal{M}_{\text{TP}}(\{q_{i,(n)}\};p'_1,p'_2)\, \mathcal{M}_{\text{TP}}^{*}(\{q_{i,(n)}\};p_1,p_2),
\label{TP_optical}
\end{align}
where both sides are restricted to $p'_1+p'_2=p_1+p_2$. Equation~\eqref{TP_optical} represents the finite-temperature generalisation of the optical theorem. The major difference with the corresponding vacuum representation is that model dependence enters both in the scattering amplitudes $\mathcal{M}_{\text{TP}}$ as well as the state normalisation factors, which depend on $\rho_{\text{TP}}(\omega,\vec{q}_{i})$. For simplicity, in this study we consider the case where Eq.~\eqref{TP_optical} is dominated by the exchange of two-thermoparticle states in the intermediate channel, i.e. $n=2$. In this case Eq.~\eqref{TP_optical} takes the form
\begin{align}
&\mathcal{M}_{\text{TP}}(p_1,p_2;p'_1,p'_2)-\mathcal{M}_{\text{TP}}^{*}(p'_1,p'_2;p_1,p_2)  \nonumber \\
&=\frac{1}{2!}\int_{-\infty}^{\infty} \frac{d \omega_{1}}{2\pi}\int \frac{d^{3} \vec{q}_{1}}{(2\pi)^{3}} \frac{\rho_{\text{TP}}(\omega_{1},\vec{q}_{1})}{1-e^{-\beta \omega_{1}}} \int_{-\infty}^{\infty} \frac{d \omega_{2}}{2\pi}\int \frac{d^{3} \vec{q}_{2}}{(2\pi)^{3}} \frac{\rho_{\text{TP}}(\omega_{2},\vec{q}_{2})}{1-e^{-\beta \omega_{2}}}  \nonumber \\
& \quad\quad\quad\quad\quad\quad\quad  \times  i(2\pi)^{4}\, \delta^{(4)}(p'_1+p'_2-q_{1}-q_{2})    \mathcal{M}_{\text{TP}}(q_{1},q_{2};p'_1,p'_2)\, \mathcal{M}_{\text{TP}}^{*}(q_{1},q_{2};p_1,p_2) \nonumber \\
&= \frac{i\pi}{2!(1-e^{-\beta E})}  \int \frac{d^{3} \vec{q}_{1}}{(2\pi)^{3}} \int_{-\infty}^{\infty} \frac{d \omega_{1}}{2\pi} \int_{-\infty}^{\infty} \frac{d \omega_{2}}{2\pi}  \, \rho_{\text{TP}}(\omega_{1},\vec{q}_{1}) \rho_{\text{TP}}(\omega_{2},\vec{P}-\vec{q}_{1})  \left(\coth\left(\tfrac{\beta\omega_{1}}{2}\right) + \coth\left(\tfrac{\beta\omega_{2}}{2}\right)   \right)  \nonumber \\
& \quad\quad\quad\quad\quad\quad\quad  \times \delta(E-\omega_{1}-\omega_{2}) \mathcal{M}_{\text{TP}}(q_{1},P-q_{1};p'_1,P-p'_1)\, \mathcal{M}_{\text{TP}}^{*}(q_{1},P-q_{1};p_1,P-p_1),
\label{TP_optical_2}
\end{align}
where in the final line the $\vec{q}_{2}$ integral is evaluated, and the implicit energy-momentum conservation restriction is used to express the integrand in terms of the total energy $E$ and momentum $\vec{P}$ of the system.

\subsection{Partial-wave projection and the two-thermoparticle phase space}
\label{subsec:partial-wave}

At zero temperature the partial-wave expansion can be regarded as a practical tool to reduce the number of kinematic variables for two scattered on-shell particles. On-shell particles fulfill the relativistic dispersion relation, which together with energy conservation fixes the magnitudes of all spatial momenta involved in the scattering process as a function of the centre-of-momentum frame energy $E^*=\sqrt{E^2-\vec{P}^2}$. Specifically, the magnitude of the momentum of any one of the particles $|\vec{q}^{\,*}|$ is fixed by $E^*$ via the relation
\begin{align}
|\vec{q}^{\,*}|=\frac{1}{2}\sqrt{E^{*}{}^{2}-4m^{2}},
\label{disp_vac}
\end{align}
and hence all remaining momentum dependence is in the angular directions. In the zero-temperature scattering amplitude this dependence can be factored out explicitly by applying the partial-wave expansion 
\begin{equation}
\mathcal{M}(p_1,P-p_1;p'_1,P-p'_1)= 4\pi \!\sum_{\ell' m',\ell m}\, Y_{\ell' m'}(\hat{p}'_1)\, \mathcal{M}_{\ell' m';\ell m}(E^*)\, Y_{\ell m}^{*}(\hat{p}_1),
\label{eq:partial-wave-exp-vacuum}
\end{equation}
where $\mathcal{M}_{\ell' m';\ell m}(E,\vec{P})$ is the partial-wave projected amplitude, $Y_{\ell m}(\hat{p}_1)$ are spherical harmonics, and $\hat{p}_1$, $\hat{p}'_1$ denote the purely angular directions. Equation~\eqref{eq:partial-wave-exp-vacuum} relies on two key properties: the existence of an on-shell dispersion relation, and the invariance under Lorentz boosts. The combination of these properties allow the explicit dependence on $\vec{P}$ to be removed. As noted previously, as soon as the system has a non-vanishing temperature both of these properties are violated, and hence any partial wave-like expansion must be qualitatively different. Although thermoparticle states do not satisfy a fixed dispersion relation, their spectral functions $\rho_{\text{TP}}(\omega,\vec{p})$ still remain peaked at some fixed energy $\omega_{\text{peak}}(|\vec{p}|,\beta)$ satisfying the condition 
\begin{align}
\left.\frac{\partial \rho_{\text{TP}}(\omega,\vec{p})}{\partial \omega}\right|_{\omega_{\text{peak}}}=0. 
\end{align}
This peak becomes increasingly pronounced at lower temperatures, and in the zero-temperature limit: $\omega_{\text{peak}}(|\vec{p}|,\beta) \rightarrow \sqrt{|\vec{p}|+m^{2}}$, as expected. $\omega_{\text{peak}}(|\vec{p}|,\beta)$ represents an effective dispersion relation whose temperature and momentum dependence is determined by the functional form of the thermoparticle spectral function. Just as the vacuum spectral function peaks imply the on-shell relation in Eq.~\eqref{disp_vac}, the thermoparticle peaks $\omega_{\text{peak}}$ will give rise to an effective two-thermoparticle dispersion relation from which one can determine the most probable magnitude of the thermoparticle momentum $|\vec{q}_{\text{TP}}|$. However, the loss of boost invariance implies that $|\vec{q}_{\text{TP}}|$ must depend separately on both $E$ and $\vec{P}$, as well as the temperature of the system. The accuracy of the $|\vec{q}_{\text{TP}}|$ approximation depends on how dominant the thermoparticle degrees of freedom are relative to the other thermal excitations, which is a purely dynamical question. Nevertheless, since $|\vec{q}_{\text{TP}}|$ smoothly approaches $|\vec{q}^{\,*}|$ in the zero-temperature limit, for sufficiently small temperatures there will exist a regime in which this serves as a good approximation. Due to the non-trivial thermal interactions $|\vec{q}_{\text{TP}}|$ can potentially have a complicated dependence on the parameters of the theory, and hence for practical calculations may need to be computed numerically. \\    

\noindent
Expanding the thermoparticle scattering amplitude $\mathcal{M}_{\text{TP}}(p_1,P-p_1;p'_1,P-p'_1)$ around the point $|\vec{p}_1|=|\vec{p}_1\!'|=|\vec{q}_{\text{TP}}|$ one can write the following effective partial-wave expansion 
\begin{align}
\mathcal{M}_{\text{TP}}(p_1,P-p_1;p'_1,P-p'_1) = 4\pi \! \sum_{\ell' m',\ell m}\, Y_{\ell' m'}(\hat{p}'_1)\, \mathcal{M}_{\ell' m';\ell m, \text{TP}}(E,\vec{P})\, Y_{\ell m}^{*}(\hat{p}_1),
\label{eq:partial-wave-th}
\end{align}
where now $\mathcal{M}_{\ell' m';\ell m, \text{TP}}(E,\vec{P})$ depends on both $E$ and $\vec{P}$ rather than only on $E^*$, as in the vacuum case. Since there is a low-temperature regime in which the magnitudes of the thermoparticle momenta are closely clustered around $|\vec{q}_{\text{TP}}|$, Eq.~\eqref{eq:partial-wave-th} will therefore provide a good approximation in this regime. At higher temperatures one could potentially include higher-order corrections in the expansion, similar to effective mass expansions in quasi-particle models. However, at some temperature this expansion will break down because the thermoparticle degrees of freedom will be significantly suppressed relative to the other thermal excitations of the system, which on the level of the spectral function implies that the thermoparticle component in Eq.~\eqref{TP_damping} no longer dominates the full thermal spectral density $\widetilde{D}_{\beta}(\vec{u},s)$ in Eq.~\eqref{rho_rep}. In the regime where Eq.~\eqref{eq:partial-wave-th} is expected to be a good approximation, the unitarity relation in Eq.~\eqref{TP_optical_2} can be formulated in terms of the partial-wave projected scattering amplitude, leading to 
\begin{align}
\text{Im} \ \mathcal{M}_{\ell ' m';\ell m, \text{TP}}(E,\vec{P})=\mathcal{M}_{\ell' m'; \ell'' m'', \text{TP}}(E,\vec{P})\,\varrho_{\ell'' m'';\ell''' m''', \text{TP}}(E,\vec{P})\, \mathcal{M}^{*}_{\ell''' m''';\ell m, \text{TP}}(E,\vec{P}),
\label{eq:optical-theorem-th}
\end{align}
where repeated indices indicate a summation, and $\varrho_{\ell m;\ell' m', \text{TP}}(E,\vec{P})$ is the thermoparticle generalisation of the zero-temperature phase space, defined as
\begin{align}
\varrho_{\ell m;\ell' m', \text{TP}}(E,\vec{P}) &= \frac{\pi^{2}}{1-e^{-\beta E}}  \int \frac{d^{3} \vec{k}}{(2\pi)^{3}} \int_{-\infty}^{\infty} \frac{d \omega_{1}}{2\pi} \int_{-\infty}^{\infty} \frac{d \omega_{2}}{2\pi}  \, \rho_{\text{TP}}(\omega_{1},\vec{k}) \rho_{\text{TP}}(\omega_{2},\vec{P}-\vec{k})  \nonumber \\
& \quad\quad\quad\quad  \times \delta(E-\omega_{1}-\omega_{2}) \left(\coth\left(\tfrac{\beta\omega_{1}}{2}\right) + \coth\left(\tfrac{\beta\omega_{2}}{2}\right)   \right) Y_{\ell m}(\hat{k}) Y_{\ell' m'}^{*}(\hat{k}).
\label{eq:phase-space-th-general-P}
\end{align}
Hereafter, we will refer to $\varrho_{\ell m;\ell' m', \text{TP}}(E,\vec{P})$ as the two-thermoparticle phase space. Since Eq.~\eqref{eq:optical-theorem-th} is a finite-temperature realisation of the optical theorem within the regime that Eq.~\eqref{eq:partial-wave-th} holds, any thermoparticle scattering amplitude that obeys Eq.~\eqref{eq:optical-theorem-th} within this regime must be expressible in the form 
\begin{align}
\mathcal{M}_{\ell'm';\ell m, \text{TP}}(E,\vec{P})=\frac{1}{\mathcal{K}^{-1}_{\ell'm';\ell m, \text{TP}}(E,\vec{P})-i\varrho_{\ell ' m';\ell m, \text{TP}}(E,\vec{P})},
\label{eq:M_K}
\end{align}
where $\mathcal{K}^{-1}_{\ell'm';\ell m, \text{TP}}(E,\vec{P})$ is the finite-temperature generalisation of the real and symmetric partial-wave-projected inverse $K$-matrix. As in the vacuum case, the singularity structure of the scattering amplitude $\mathcal{M}_{\ell'm';\ell m, \text{TP}}(E,\vec{P})$ provides information about the states that exist in the theory. However, the presence of temperature introduces additional subtleties which need to be taken into account in order to establish the true nature of these states, in particular whether they represent bound-state or resonance-like excitations.

\subsection{Two-thermoparticle phase space in the rest frame} 
\label{subsec:rest-frame-phase-space}

For simplicity we now focus on the calculation of $\varrho_{\ell m;\ell' m', \text{TP}}(E,\vec{P})$ for $\vec{P}=0$, which corresponds to the rest frame of the medium, or equivalently the thermoparticle centre-of-momentum frame. From Eq.~\eqref{eq:phase-space-th-general-P} it follows that $\varrho_{\ell m;\ell' m', \text{TP}}(E):=\varrho_{\ell m;\ell' m', \text{TP}}(E,\vec{P}=0)$ can be written  
\begin{align}
\varrho_{\ell m;\ell' m', \text{TP}}(E) &= \frac{1}{8\pi(1-e^{-\beta E})} \, \delta_{\ell \ell'}\,\delta_{mm'} \int_{0}^{\infty}\! d|\vec{k}| \,|\vec{k}|^{2}  \int_{-\infty}^{\infty} \frac{d \omega_{1}}{2\pi} \int_{-\infty}^{\infty} \frac{d \omega_{2}}{2\pi}  \, \rho_{\text{TP}}(\omega_{1},\vec{k}) \rho_{\text{TP}}(\omega_{2},\vec{k})  \nonumber \\
& \quad\quad\quad\quad\quad\quad\quad\quad\quad\quad\quad\quad  \times \delta(E-\omega_{1}-\omega_{2}) \left(\coth\left(\tfrac{\beta\omega_{1}}{2}\right) + \coth\left(\tfrac{\beta\omega_{2}}{2}\right)   \right),
\label{eq:phase-space-rest}
\end{align}
which follows from the angular-independence of the spectral functions and the orthonormality of the spherical harmonics. The explicit factor of $\delta_{\ell \ell'}\,\delta_{mm'}$ demonstrates that the phase space is diagonal in the angular momentum variables, as required by angular momentum conservation. Due to the threshold characteristics of the thermoparticle spectral function in Eq.~\eqref{TP_thresh}, only integration regions with $\omega_{i}^{2} \geq m^{2}$ will contribute in Eq.~\eqref{eq:phase-space-rest}. Taking each of these integration regions into account, using the anti-symmetry of the spectral function $\rho_{\text{TP}}(-\omega_{i},\vec{k})=-\rho_{\text{TP}}(\omega_{i},\vec{k})$, and performing the $\omega_{2}$ integral, Eq.~\eqref{eq:phase-space-rest} takes the form 
\begin{align}
\varrho_{\ell m;\ell' m', \text{TP}}(E) &= \frac{1}{8\pi^{2}(1-e^{-\beta E})} \, \delta_{\ell \ell'}\,\delta_{mm'} \int_{0}^{\infty}\! d|\vec{k}| \,|\vec{k}|^{2}\int_{m}^{\infty}\frac{d \omega_{1}}{2\pi}  \, \rho_{\text{TP}}(\omega_{1},\vec{k}) \coth\left(\tfrac{\beta\omega_{1}}{2}\right)    \nonumber \\
& \quad \times  \left[  \rho_{\text{TP}}(E - \omega_{1},\vec{k})    \theta(E-\omega_{1}-m)  -\rho_{\text{TP}}(\omega_{1}-E,\vec{k})  \theta(\omega_{1}-E -m)  \right. \nonumber \\ 
&\left. \quad\quad +\rho_{\text{TP}}(E+\omega_{1},\vec{k})  \theta(E + \omega_{1}-m) - \rho_{\text{TP}}(-E - \omega_{1},\vec{k})    \theta(-E-\omega_{1}-m)  \right].
\label{eq:phase-space-rest-2}
\end{align}
Using the fact that the integrand is anti-symmetric under the interchange $E \rightarrow -E$, together with the restriction on the integration regions from the theta terms, $\varrho_{\ell m;\ell' m', \text{TP}}(E)$ can finally be written  
\begin{align}
\varrho_{\ell m;\ell' m', \text{TP}}(E) &= \frac{\epsilon(E)}{8\pi^{2}(1-e^{-\beta E})} \, \delta_{\ell \ell'}\,\delta_{mm'} \int_{0}^{\infty}\! d|\vec{k}| \,|\vec{k}|^{2}  \nonumber \\
& \quad\quad \times \left[  \theta(|E|-2m) \int_{m}^{|E|-m} \frac{d\omega_{1}}{2\pi}   \, \rho_{\text{TP}}(\omega_{1},\vec{k}) \rho_{\text{TP}}(|E|-\omega_{1},\vec{k})\coth\left(\tfrac{\beta\omega_{1}}{2}\right)     \right. \nonumber \\ 
& \quad\quad\quad\quad  + \left( \int_{m}^{\infty} \frac{d\omega_{1}}{2\pi}  \, \rho_{\text{TP}}(\omega_{1},\vec{k}) \rho_{\text{TP}}(|E|+\omega_{1},\vec{k}) \right. \nonumber \\
& \quad\quad\quad\quad\quad\quad\quad\quad  \left. \left. - \int_{|E|+m}^{\infty} \frac{d\omega_{1}}{2\pi}  \, \rho_{\text{TP}}(\omega_{1},\vec{k}) \rho_{\text{TP}}(\omega_{1}-|E|,\vec{k})  \right)\coth\left(\tfrac{\beta\omega_{1}}{2}\right)  \right] \!,
\label{eq:phase-vacuum-landau-split}
\end{align}
where $\epsilon(E)$ is the sign function. From Eq.~\eqref{eq:phase-vacuum-landau-split} one can see that $\varrho_{\ell m;\ell' m', \text{TP}}(E)$ contains both threshold ($|E| \geq 2m$) and non-threshold contributions. In the zero-temperature limit: $\rho_{\text{TP}}(\omega_{1},\vec{k}) \rightarrow 2\pi \epsilon(\omega_{1})\delta(\omega_{1}^{2}-|\vec{k}|^{2}-m^{2})$, and the non-threshold components exactly vanish. These components therefore represent purely thermal excitations, with non-vanishing spectral contributions both above and below the vacuum particle threshold $E=2m$. As discussed in Sec.~\ref{sec:TP_basics}, finite-temperature systems possess both positive and negative energy states, with the latter corresponding to the potential to extract energy from the thermal medium itself. This property is reflected in the fact that $\varrho_{\ell m;\ell' m', \text{TP}}(E)$ is non-vanishing for $E<0$. However, these excitations are increasingly damped for larger values of the inverse temperature $\beta$ by the explicit $(1-e^{-\beta E})^{-1}$ factor in Eq.~\eqref{eq:phase-vacuum-landau-split}, which represents their thermodynamic suppression. For the purely threshold term one finds that in the zero-temperature limit this reduces to the standard vacuum two-particle phase space
\begin{align}
\varrho_{\ell m;\ell' m'}(E)=\delta_{ll'}\, \delta_{mm'}\,\frac{\sqrt{E^2-4m^2}}{32\pi E}\,\theta(E-2m),
\label{eq:zero-T-phase-space}
\end{align}
where $E=E^{*}$ in this case since $\vec{P}=0$, and the negative-energy components are no longer present. That the vacuum threshold component $\theta(E-2m)$ still persists at finite-temperature follows from Eq.~\eqref{TP_thresh}, which encodes the fact that the medium must be supplied with an energy of at least the vacuum particle mass $m$ in order to create a single-thermoparticle state with non-negligible probability. By considering non-trivial thermal scattering states the phase space in Eq.~\eqref{eq:phase-vacuum-landau-split} has a fundamentally different structure to previous results in the literature~\cite{GomezNicola:2023rqi}, which have relied on standard perturbative or chiral effective field theory techniques. Since the scattering states in these approaches are non-interacting, this gives rise to a phase space which is proportional to the zero-temperature result. However, this neglects the presence of purely thermal excitations that can have below threshold energies, as well as the possibility of negative energy hole-like states, which must exist by virtue of the KMS condition, as discussed in Sec.~\ref{sec:TP_basics}. \\

\noindent
Since the properties of thermoparticles are determined by the dynamics of the theory, the specific structure of the two-thermoparticle phase space $\varrho_{\ell m;\ell' m', \text{TP}}(E)$ depends on the damping experienced by these states. Therefore, in contrast to the vacuum case, $\varrho_{\ell m;\ell' m', \text{TP}}(E)$ is model dependent. To establish the form of $\varrho_{\ell m;\ell' m', \text{TP}}(E)$ one needs to know the structure of the damping factor $D_{m,\beta}(\vec{x})$, since this determines the corresponding thermoparticle spectral function $\rho_{\text{TP}}(\omega,\vec{p})$. In Ref.~\cite{Bros:2001zs} the authors established a consistency condition for thermal asymptotic scattering states, and found that this uniquely fixes the form of $D_{m,\beta}(\vec{x})$, demonstrating that the structure of the thermoparticle states can in principle be computed for any QFT from the dynamical equations of the theory. In practice, it turns out that one can also extract the form of the damping factors directly from two-point correlation function data. Recently, this approach has been successfully applied to lattice data in real~\cite{Lowdon:2024atn,Ali:2026ehk} and complex~\cite{Lowdon:2025fyb,Lowdon:2025ait} scalar theories, as well as QCD~\cite{Lowdon:2022xcl,Bala:2023iqu}. \\

\begin{figure}[t!]
\centering
\includegraphics[width=0.49\textwidth]{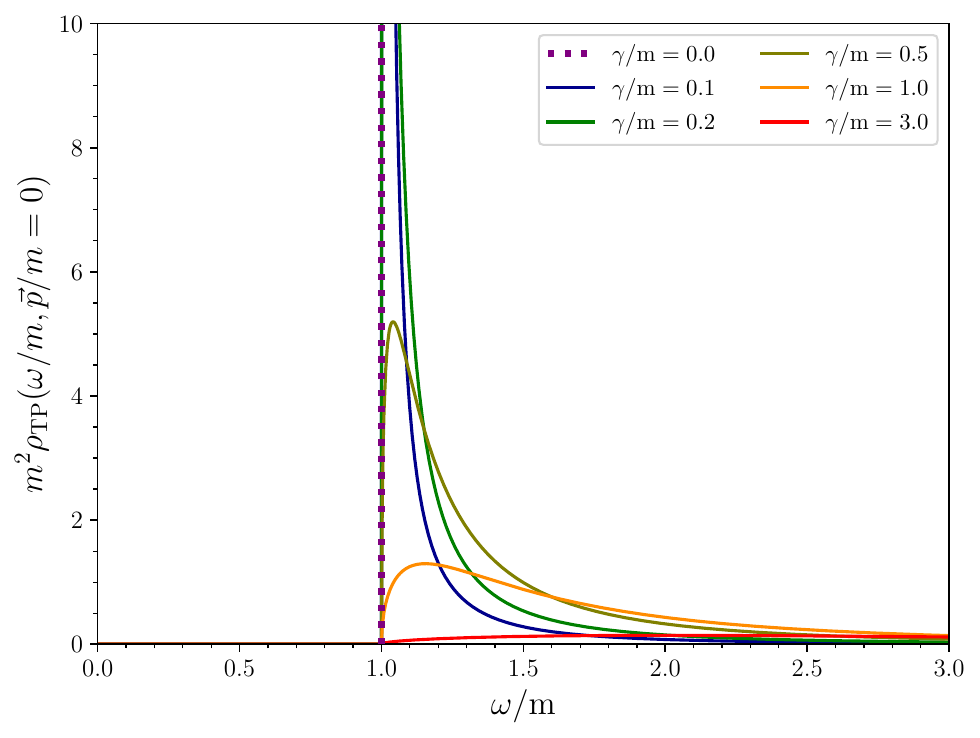}
\includegraphics[width=0.49\textwidth]{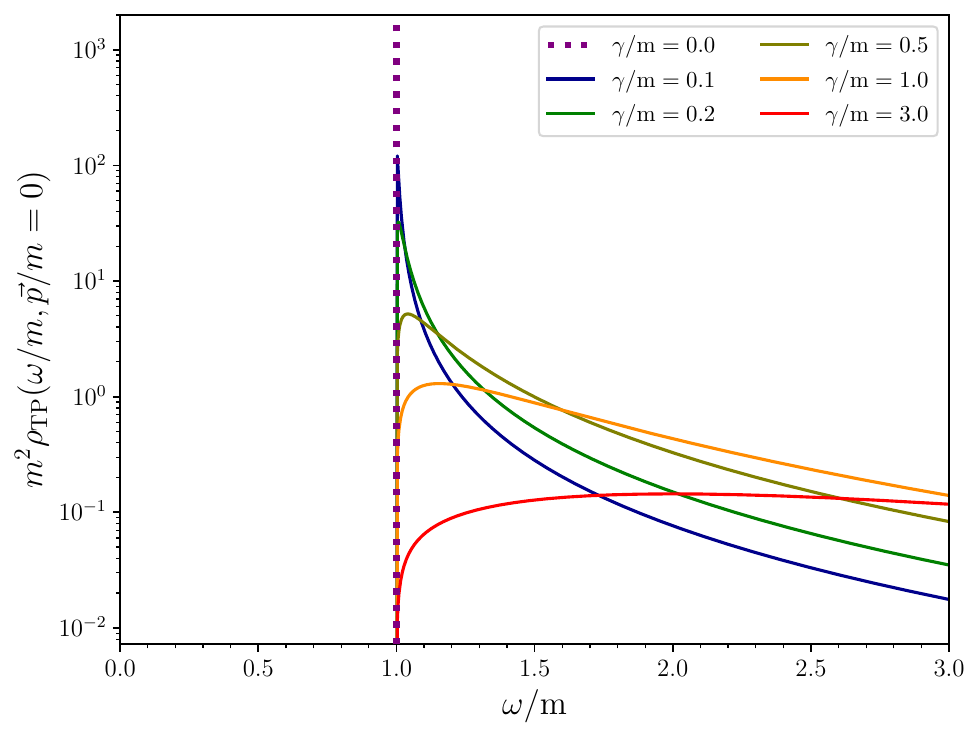}
\caption{The thermoparticle spectral function $\rho_{\text{TP}}(\omega,\vec{p}=0)$ for an exponential damping factor $D_{m,\beta}=e^{-\gamma |\vec{x}|}$ at $\gamma/m =$ 0, 0.1, 0.2, 0.5, 1.0, 3.0 plotted on a non-logarithmic (left) and logarithmic (right) scale.}
\label{fig:TP_spectral}
\end{figure} 

\noindent
In order to illustrate the differences between $\varrho_{\ell m;\ell' m', \text{TP}}(E)$ and the vacuum phase space in Eq.~\eqref{eq:zero-T-phase-space}, we chose the damping factor to have the purely exponential form
\begin{align}
D_{m,\beta}(\vec{x}) = \alpha \, e^{-\gamma |\vec{x}|},
\label{exp_damping}
\end{align} 
as seen in lattice $\phi^{4}$ theory~\cite{Lowdon:2024atn,Ali:2026ehk}. With the spectral representation in Eq.~\eqref{rho_rep} it follows that
\begin{align}
\rho_{\text{TP}}(\omega,\vec{p}) &= \epsilon(\omega)  \theta(\omega^{2}-m^{2}) \,  \frac{4 \alpha \gamma  \sqrt{\omega^{2}-m^{2}}}{(|\vec{p}|^{2}+m^{2}-\omega^{2})^{2} + 2(|\vec{p}|^{2}-m^{2}+\omega^{2})\gamma^{2}+\gamma^{4} }, \label{rho_TP_0}  
\end{align}
where $\alpha$ and $\gamma$ are temperature-dependent parameters. $\gamma$ represents a thermal width-like parameter which also depends on the coupling of the system, but for simplicity we set $\gamma=T$ and fix $\alpha=1$. In Fig.~\ref{fig:TP_spectral} the thermoparticle spectral function in Eq.~\eqref{rho_TP_0} is plotted for $\vec{p}=0$ at different values of $\gamma/m$. For $\gamma/m>0$ the spectral function has a finite peak, which broadens for increasing temperature. In the zero-temperature limit the thermal width $\gamma$ vanishes, and Eq.~\eqref{rho_TP_0} reduces to the spectral function $\rho_{\text{vac}}(\omega,\vec{p})=2\pi \alpha \,\epsilon(\omega)\delta(p^{2}-m^{2})$ of an on-shell particle of mass $m$. To compute the integrals in Eq.~\eqref{eq:phase-vacuum-landau-split} we used the multi-dimensional integration package \textit{cubature}~\cite{cubature_c}. For the upper limits of the $\omega_1$ and $|\vec{k}|$ integration regions we chose large but finite cutoffs $(\Lambda_{\omega_1}=100m,\Lambda_k=1000m)$, which ensured that the integration result was insensitive to their variation. We used a higher number of integration steps to compute the non-threshold components in Eq.~\eqref{eq:phase-vacuum-landau-split}, since these components led to significantly more numerical noise than the threshold integral. \\ 

\begin{figure}[t!]
\centering
\includegraphics[width=0.49\textwidth]{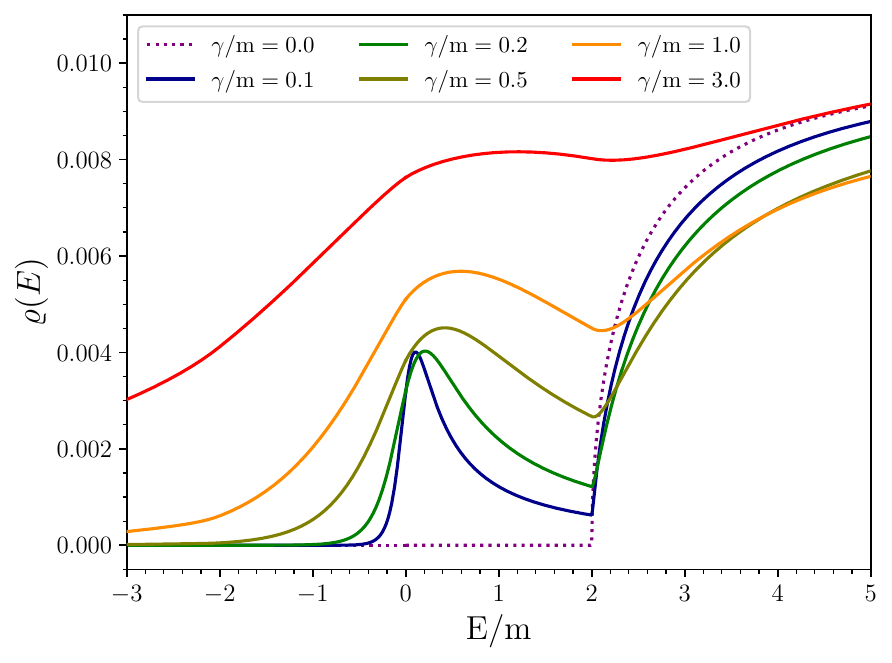}
\includegraphics[width=0.49\textwidth]{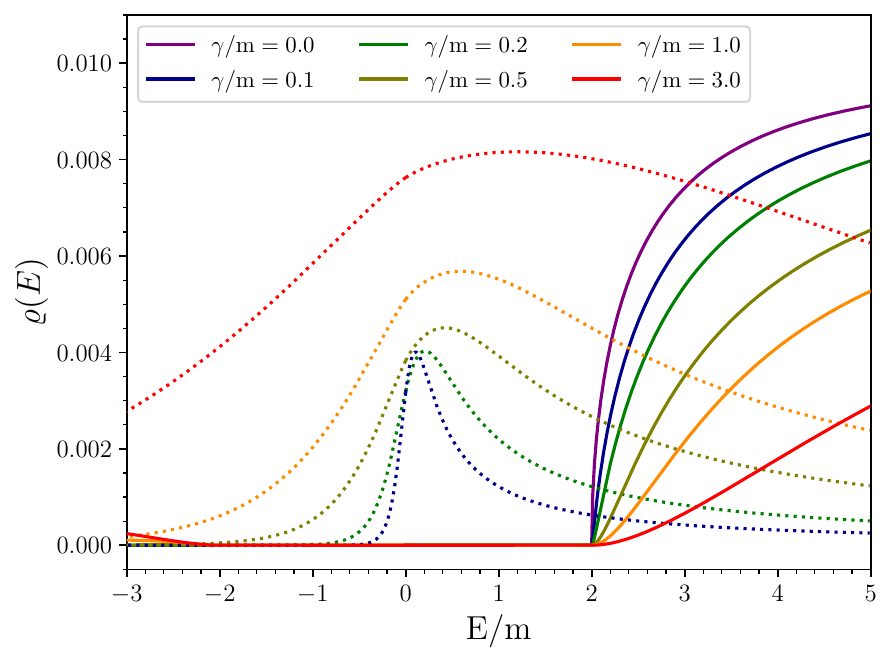}
\caption{Two-thermoparticle phase space $\varrho(E) := \varrho_{\ell m;\ell m, \text{TP}}(E)$ with an exponential damping factor $D_{m,\beta}(\vec{x}) = e^{-\gamma |\vec{x}|}$, at $\gamma/m =$ 0, 0.1, 0.2, 0.5, 1.0, 3.0. The left plot shows the total phase space, and the right plot shows the separate threshold (solid lines) and non-threshold (dotted lines) components.}
\label{fig:full-phase-space}
\end{figure}

\noindent
In Fig.~\ref{fig:full-phase-space} $\varrho_{\ell m;\ell m, \text{TP}}(E)$ is plotted in the region $-3 \leq E/m \leq 5$ using the same range of thermal widths $\gamma/m\in \{0.0,0.2,0.5,1.0,3.0\}$ as for the thermoparticle spectral function in Fig.~\ref{fig:TP_spectral}. The left plot shows the total phase space, and in the right plot the threshold (solid lines) and non-threshold (dotted lines) components are shown separately. For the vacuum case ($\gamma=0$) the non-threshold component exactly vanishes, and the phase space has the form in Eq.~\eqref{eq:zero-T-phase-space}, with a branch point singularity at the two-particle threshold $E=2m$. As the temperature and hence $\gamma$ increases the non-threshold component becomes non-vanishing and increases in amplitude both above and below $E=2m$, whilst the threshold component is suppressed. As a result, the vacuum threshold becomes increasingly screened, to the point where its signature is effectively absent in the total phase space, as seen for $\gamma/m=3.0$. Physically, this represents the point at which the thermoparticles are no longer the dominant degrees of freedom, since they are overwhelmed by the purely thermal excitations of the medium. In Fig.~\ref{fig:full-phase-space} one can also see that there are non-vanishing negative energy contributions, but these are highly suppressed for lower temperatures, as discussed in Sec.~\ref{subsec:rest-frame-phase-space}. Interestingly, at $E=0$ the phase space is non-zero, in contrast to the vacuum case. The excitations at this point represent thermoparticle-hole-like states $|E\rangle \otimes |-E\rangle$ with vanishing total energy. The thermal phase space displayed in Fig.~\ref{fig:full-phase-space} is significantly different to the standard propositions in the literature~\cite{GomezNicola:2023rqi}, where only above threshold $E > 2m$ components exist at all temperatures. These differences stem from the inclusion of interaction effects in the thermal scattering states, which give rise to purely thermal non-threshold components, as well as the potential to create negative-energy states.

\section{Two-particle quantisation condition at finite temperature} 
\label{sec:qc2-derivation} 

\subsection{Zero-temperature condition}
\label{subsec:zero-temp}

The notion of asymptotic states, and hence observables such as the two-particle scattering amplitude, are only well-defined in an infinite spatial volume, whereas lattice calculations are performed in a finite volume. Nevertheless, information about physical scattering amplitudes \textit{can} be extracted from lattice calculations. In particular, finite-volume lattice energy levels $\mathcal{E}(\vec{P};L)$ are connected to the infinite-volume two-particle scattering amplitude $\mathcal{M}(E)$ via the \textit{two-particle quantisation condition}, which has the form~\cite{Luscher:1986Scat,Luscher:1991ux,Kim:2005gf,Hansen:2012tf}
\begin{align}
\text{det}\,[F^{-1}(E,\vec{P};L) + \mathcal{M}(E)]=0,
\label{eq:vacuum-quantisation-condition}
\end{align}
where $F^{-1}$ encodes information about the finite-volume effects of the system, and both $F^{-1}$ and $\mathcal{M}$ are matrices in angular momentum space. Solving Eq.~\eqref{eq:vacuum-quantisation-condition} for $E=\mathcal{E}(\vec{P};L)$ constrains the form of the scattering amplitude $\mathcal{M}$, and provides a first principle approach by which one can infer information about the properties of resonance-like states such as the $\rho$ meson in QCD, which are intrinsically unstable in vacuum. An implicit assumption in the derivation of Eq.~\eqref{eq:vacuum-quantisation-condition} is that the spatial lattice volume $L^{3}$ is finite, but the temporal extent $L_{\tau}$ is infinite. Since the temperature of a lattice-discretised system is defined as $T=1/L_{\tau}$~\cite{Montvay:1994cy}, this is equivalent to the assumption that the system is at zero temperature. However, since $L_{\tau}= \beta$ is finite in any practical lattice simulation, it is important to understand the effect that this has on the quantisation condition. Not only can this help in quantifying the systematic uncertainties from finite-$L_{\tau}$ effects in the extraction of vacuum observables, but this also opens the possibility of studying the in-medium effects experienced by resonance states such as the $\rho$ meson at finite temperature. Since thermoparticles provide a consistent description of thermal scattering states, as discussed in Secs.~\ref{sec:TP_basics} and~\ref{sec:infinite-volume-scattering}, it is natural that the finite-temperature generalisation of Eq.~\eqref{eq:vacuum-quantisation-condition} is parametrised in terms of these degrees of freedom. In this section we will derive this generalisation for two thermoparticles on a finite lattice $L_{\tau} \times L^{3}$, and explore its consequences for the specific thermoparticle parametrisation in Sec.~\ref{subsec:rest-frame-phase-space}.

\subsection{Finite-volume thermal two-point function}

The derivation of the finite-temperature generalisation of Eq.~\eqref{eq:vacuum-quantisation-condition} closely follows the strategy developed in Refs.~\cite{Kim:2005gf,Hansen:2012tf}. We consider the case of two identical scalar thermoparticle states, whose corresponding spectral functions $\rho_{\text{TP}}(\omega,\vec{p})$ approach those of a stable vacuum particle state of mass $m$ in the zero-temperature limit. In Euclidean spacetime, the presence of a non-vanishing temperature $T=1/\beta>0$ means that the thermal two-point correlation function
\begin{align}
C(\tau,\vec{x}) = \langle \mathcal{O}(\tau,\vec{x})\, \mathcal{O}^{\dagger}(0,\vec{0})\rangle_{\beta},
\label{corrT}
\end{align}
is $\beta$-periodic in imaginary time $\tau$. This follows from the KMS condition in Eq.~\eqref{KMS}. On a finite $L_{\tau} \times L^{3}$ lattice this is equivalent to choosing a compactified $\tau$ direction with periodic boundary conditions, and implies that the Euclidean energy variable $k_{4}$, which is conjugate to $\tau$, must take discrete values
\begin{align}
k_{4}=\omega_{n_{4}}=\frac{2\pi n_4}{\beta}, \quad n_4\in\mathbb{Z},
\end{align}
where $\omega_{n_{4}}$ are the Matsubara frequencies. By choosing periodic boundary conditions also in the spatial directions, this similarly implies the discretisation of the spatial momenta in the system: $\vec{k}=\tfrac{2\pi \vec{n}}{L}, \, \vec{n} \in \mathbb{Z}^{3}$. The key difference compared to the zero-temperature scenario studied in Refs.~\cite{Kim:2005gf,Hansen:2012tf} is that the Euclidean energies $k_4$ are no longer continuous, and hence any continuous $k_4$ integral must be replaced with a Matsubara sum:
\begin{align}
\int \frac{dk_4}{2\pi} \rightarrow \frac{1}{\beta} \! \sum_{k_4=\omega_{n_{4}}}
\label{eq:Matsubara}
\end{align} 
As is already central to the derivations in Refs.~\cite{Kim:2005gf,Hansen:2012tf}, taking into account finite spatial volume effects requires the replacement: $\int \!\frac{d^3 k}{(2\pi)^3}\rightarrow \tfrac{1}{L^3}\,\sum_{\vec{k}\in \frac{2\pi}{L}\mathbb{Z}^3}$. In general, for any finite lattice the spacetime points are restricted to a subset of $a\mathbb{Z}^{4}$, where $a>0$ is the lattice spacing. This means that any momentum-space function is $2\pi/a$-periodic, and hence the Euclidean energy and spatial momenta are restricted to the first Brillouin zone: $-\tfrac{\pi}{a} < k_{\mu} \leq \tfrac{\pi}{a}$, resulting in the $\vec{k}$ and Matsubara sum in Eq.~\eqref{eq:Matsubara} being finite. However, as in Refs.~\cite{Kim:2005gf,Hansen:2012tf}, for simplicity we will ignore these effects\footnote{See Ref.~\cite{Hansen:2024cai} for a recent discussion of discretisation effects in the two-particle quantisation condition.} and consider the continuum limit $a\rightarrow 0$.  \\

\noindent
On the $L_{\tau} \times L^{3}$ lattice the momentum-projected Euclidean two-point correlator has the form  
\begin{align}
C_{L,\beta}(\tau,\vec{P})= \int_{L} d^{3}\vec{x} \int_{L} d^{3}\vec{y} \ e^{-i \vec{P}\cdot(\vec{x}-\vec{y})} \langle \mathcal{O}(\tau,\vec{x})\, \mathcal{O}^{\dagger}(0,\vec{y})\rangle_{\beta} =  \langle \mathcal{O}(\tau,\vec{P})\, \mathcal{O}^{\dagger}(0,-\vec{P})\rangle_{\beta},
\label{eq:finite-volume-correlator-tau}
\end{align}
where for the purpose of this analysis we take $\mathcal{O}^{\dagger}(0,-\vec{P})$, $\mathcal{O}(\tau,\vec{P})$ to be suitable two-particle interpolating operators that respectively create a two-particle thermal state at $\tau=0$, and annihilate it at $\tau$ with fixed total momentum $\vec{P}$. The spatial integrals in Eq.~\eqref{eq:finite-volume-correlator-tau} are performed over the finite volume $L^{3}$, and the thermal expectation value is defined as\footnote{This standard finite-volume definition of the thermal expectation value can be proven to follow from the KMS condition in Eq.~\eqref{KMS}, which holds for both finite and infinite-volume systems~\cite{Haag:1992hx}.}
\begin{align}
\langle \mathcal{O}(\tau,\vec{P})\, \mathcal{O}^{\dagger}(0,-\vec{P})\rangle_{\beta}=\frac{1}{Z}\, \sum_{n}\, \langle n| \mathcal{O}(\tau,\vec{P})\, \mathcal{O}^{\dagger}(0,-\vec{P})| n \rangle\, e^{-\beta E_n},
\label{KMS_corr}
\end{align}
where $E_n$ are the discrete eigenvalues of the finite-volume Hamiltonian in the vacuum theory, and $Z$ is the partition function. $C_{L,\beta}(\tau,\vec{P})$ contains information about the thermal spectrum of the theory, in particular the energy levels $E(\vec{P};L,\beta)$ of the finite-volume states, which can be directly extracted from lattice data. Taking the Fourier transform in the compact $\tau$ direction gives the full Euclidean momentum-space correlator  
\begin{align}
C_{L,\beta}(P_4,\vec{P})=\int_{0}^{\beta} d\tau \, e^{iP_4\tau} \langle \mathcal{O}(\tau,\vec{P})\, \mathcal{O}^{\dagger}(0,-\vec{P})\rangle_{\beta}.
\label{eq:finite-volume-correlator-fourier}
\end{align}
$C_{L,\beta}(P_4,\vec{P})$ is the central quantity of interest in this analysis. At zero temperature its $L$ dependence gives rise to the two-particle quantisation condition in Eq.~\eqref{eq:vacuum-quantisation-condition}, and for $\beta=L_{\tau}>0$ its dependence on $\beta$ determines how this condition is modified by thermal effects. As always, one must perform the analytic continuation $P_4 \rightarrow -iE$ in order to recover information at physical energies $E$. \\  
 
\begin{figure}[t!]
\centering
\includegraphics[width=0.75\textwidth]{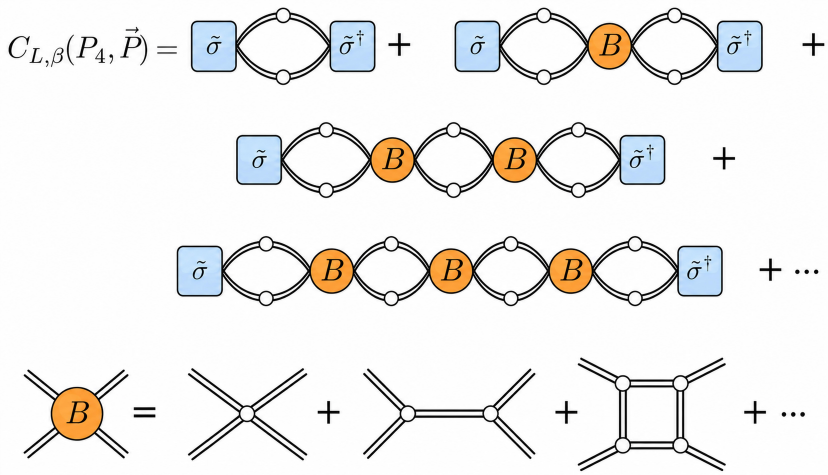}
\caption{Thermoparticle skeleton expansion of the finite-volume correlation function $C_{L,\beta}(P_4,\vec{P})$. The double lines represent thermoparticle propagators, the white dots indicate finite-volume momentum and energy sums, and the intermediate circles are thermal Bethe-Salpeter kernels.}
\label{fig:RFT-expansion}
\end{figure}

\noindent
The idea behind the approach developed in Refs.~\cite{Kim:2005gf,Hansen:2012tf} is to perform a skeleton expansion of the finite-volume correlation function, and to identify the singularity structure of the different skeleton diagrams appearing in the expansion. This corresponds to the $T=0$ analogue of Fig.~\ref{fig:RFT-expansion}. In the zero-temperature formalism the lines in this expansion represent fully-dressed single-particle vacuum propagators $\widetilde{G}_{\text{vac}}(k_{0},\vec{p})$. Analysing the $L$ dependence of these propagators one finds that the single-particle pole part of the propagator has a power-like $L^{-n}$ scaling behaviour, whereas the contributions arising from continuous spectral components decay exponentially~\cite{Kim:2005gf,Hansen:2012tf}. By ignoring the sub-leading volume corrections the skeleton expansion of $C_{L,\infty}(P_4,\vec{P})$ is therefore dominated by diagrams where the lines involve only the pole part of the propagators. At finite temperature the situation is more complicated, since the singularity structure of the thermal correlators is model dependent. Nevertheless, as outlined in Secs.~\ref{sec:TP_basics} and~\ref{sec:infinite-volume-scattering}, there is both theoretical and numerical lattice evidence that thermoparticles dominate the low-energy behaviour of correlation functions at low temperatures, i.e. large $\beta$. So although thermoparticle contributions to the spectral function $\rho_{\text{TP}}(\omega,\vec{p})$ can in principle give rise to more general types of propagator singularities, such as branch points, the contribution from these singularities will be less suppressed at large $\beta$ than the other components appearing in the full spectral function, regardless of their large-$L$ scaling. The physical picture here is that there is a low-temperature regime in which the thermal manifestation of the vacuum particle states, the thermoparticles, still play a dominant role because their spectral function peaks have not yet become sufficiently broadened by in-medium effects. Lattice analyses of both scalar theories~\cite{Lowdon:2024atn,Ali:2026ehk,Lowdon:2025fyb,Lowdon:2025ait} and QCD~\cite{Lowdon:2022xcl,Bala:2023iqu} show that this is indeed the case for $T \sim m$, where $m$ is the vacuum mass of the system. \\

\noindent
With this thermoparticle dominance in mind one can therefore generalise the vacuum approach of Refs.~\cite{Kim:2005gf,Hansen:2012tf} to low-temperature (large $\beta$) regimes by performing the replacement
\begin{align}
\widetilde{G}_{\text{vac}}(k_{0},\vec{p}) \rightarrow \widetilde{G}_{\text{TP}}(k_{0},\vec{p}),
\end{align}
where $\widetilde{G}_{\text{TP}}(k_{0},\vec{p})$ is the corresponding thermoparticle propagator defined in Eq.~\eqref{G_TP}. In the zero-temperature limit $\widetilde{G}_{\text{TP}}(k_{0},\vec{p})$ reduces to the purely pole part of $\widetilde{G}_{\text{vac}}(k_{0},\vec{p})$ with vacuum mass $m$, and one recovers the results of the standard approach. Because the thermoparticle spectral function $\rho_{\text{TP}}(\omega,\vec{p})$ vanishes for $\omega<m$, for energies $m<E<3m$ the leading finite-volume corrections are generated by combinations of two thermoparticle propagators, since all other spectral contributions that these propagators generate can be considered smooth in the region $E<3m$, and therefore receive an enhanced $e^{-mL}$ suppression. This establishes a finite-temperature generalisation of the hierarchy used in Ref.~\cite{Hansen:2012tf}. Due to this separation scheme, one can divide all possible diagrams that appear in the general skeleton expansion of $C_{L,\beta}(P_4,\vec{P})$ into two classes: two-thermoparticle loops, and thermal Bethe-Salpeter kernels. The two-thermoparticle loops contain finite-volume effects from the thermoparticle propagators. All remaining interactions are formally collected into the finite-volume thermal Bethe-Salpeter kernels $B_{L,\beta}(k'_1,k'_2;k_1,k_2)$, where $k_1, k_2$ and $k_1', k_2'$ denote the ingoing and outgoing four-momenta of the two thermoparticles. Following this expansion, $C_{L,\beta}(P_4,\vec{P})$ can be written in the form  
\begin{align}
C_{L,\beta}(P_4,\vec{P})=C_{L,\beta}^{(1)}(P_4,\vec{P})+\sum_{n=2}^{\infty} \tilde{\mathcal{\sigma}}\circ[B_{L,\beta}\,\circ]^{n-1} \tilde{\mathcal{\sigma}}^{\dagger},
\label{eq:skeleton-expansion-finite-kernels}
\end{align}
where $C_{L,\beta}^{(1)}(P_4,\vec{P})$ denotes the contribution that contains a single finite-volume two-thermoparticle loop, but no Bethe-Salpeter kernels. $\tilde{\sigma}$ and $\tilde{\sigma}^{\dagger}$ are endcap functions defined by the Euclidean Fourier transform of the position-space operator $\mathcal{O}(x,y)$ coupled to the two-thermoparticle state $|P,k\rangle_{\text{TP}}$ 
\begin{align}
\tilde{\sigma}_{\alpha}(P,k) &= \int_{0}^{\beta} dx_4 \int_{0}^{\beta} dy_4 \int d^3\vec{x}  \int d^3\vec{y} \  {}_{\beta} \langle \alpha |  \mathcal{O}(x,y) |P,k\rangle_{\text{TP}} \ e^{ikx}\, e^{i(P-k)y}, \\
\tilde{\sigma}_{\alpha}^{\dagger}(P,k) &= \int_{0}^{\beta} dx_4 \int_{0}^{\beta} dy_4 \int d^3\vec{x} \int d^3\vec{y} \  {}_{\text{TP}}\langle P,k| \mathcal{O}^{\dagger}(x,y) |\alpha\rangle_{\beta} \ e^{-ik\cdot x}\, e^{-i(P-k)\cdot y},
\end{align} 
where $|\alpha\rangle_{\beta}$ denotes a general thermal state with some specified quantum numbers $\alpha$. The operation $\circ$ in Eq.~\eqref{eq:skeleton-expansion-finite-kernels} represents a two-thermoparticle loop insertion. Given generic kinematic functions $\mathcal{L}$ and $\mathcal{R}$ to the left and right of the two-thermoparticle loop, this operation is defined 
\begin{align}
\mathcal{L}\circ \mathcal{R} = \frac{1}{2!} \, \frac{1}{L^3} \!\sum_{\vec{k}\in \frac{2\pi}{L}\,\mathbb{Z}^3} \frac{1}{\beta} \!\sum_{k_4=\omega_n} \! \mathcal{L}(P,k)\, \widetilde{G}_{\text{TP}}(k)\, \widetilde{G}_{\text{TP}}(P-k)\, \mathcal{R}(P,k),
\label{eq:L-R-Product}
\end{align}
where the $1/2!$ coefficient is a symmetrisation factor analogous to that introduced in Eq.~\eqref{TP_optical_2}. In the above equation all renormalisation factors of the thermoparticle propagators are absorbed into the functions $\mathcal{L}$ and $\mathcal{R}$, which represent either endcap functions or Bethe-Salpeter kernels. More than one application of the operation $\circ$ corresponds to nested insertions of two-thermoparticle loops. A diagrammatic representation of the decomposition of $C_{L,\beta}(P_4,\vec{P})$ in Eq.~\eqref{eq:skeleton-expansion-finite-kernels} is displayed in the upper panel of Fig.~\ref{fig:RFT-expansion}. To draw a distinction with the zero-temperature case, the thermoparticle propagators $\widetilde{G}_{\text{TP}}$ are displayed as double lines, where the internal loops involve finite-volume momentum and energy sums. The Bethe-Salpeter kernels are indicated by circles between the two-thermoparticle loops, and the leading contributions to these kernels are shown in the lower panel of Fig.~\ref{fig:RFT-expansion}. 

\subsection{Finite-volume two-thermoparticle loop and Bethe-Salpeter kernel}

Having defined all the essential components in the skeleton expansion of the finite-volume thermal correlator, namely the two-thermoparticle loop $\mathcal{L}\circ \mathcal{R}$ and Bethe-Salpeter kernel $B_{L,\beta}(k'_1,k'_2;k_1,k_2)$, we next turn to the analysis of their finite-volume effects. In the two-thermoparticle loop in Eq.~\eqref{eq:L-R-Product} one can keep the dependence on the external momentum $P$ of the functions $\mathcal{L}$ and $\mathcal{R}$ implicit and write
\begin{align}
\mathcal{L}\circ \mathcal{R} = \frac{1}{2!} \, \frac{1}{L^3} \!\sum_{\vec{k}\in \frac{2\pi}{L}\,\mathbb{Z}^3} \frac{1}{\beta} \!\sum_{k_4=\omega_n} \! \mathcal{L}(k_4,\vec{k})\, \widetilde{G}_{\text{TP}}(k_4,\vec{k})\, \widetilde{G}_{\text{TP}}(P_4-k_4,\vec{P}-\vec{k})\, \mathcal{R}(k_4,\vec{k}).
\label{eq:L-R-Product_2}
\end{align} 

\begin{figure}[t!]
\centering
\includegraphics[width=0.6\textwidth]{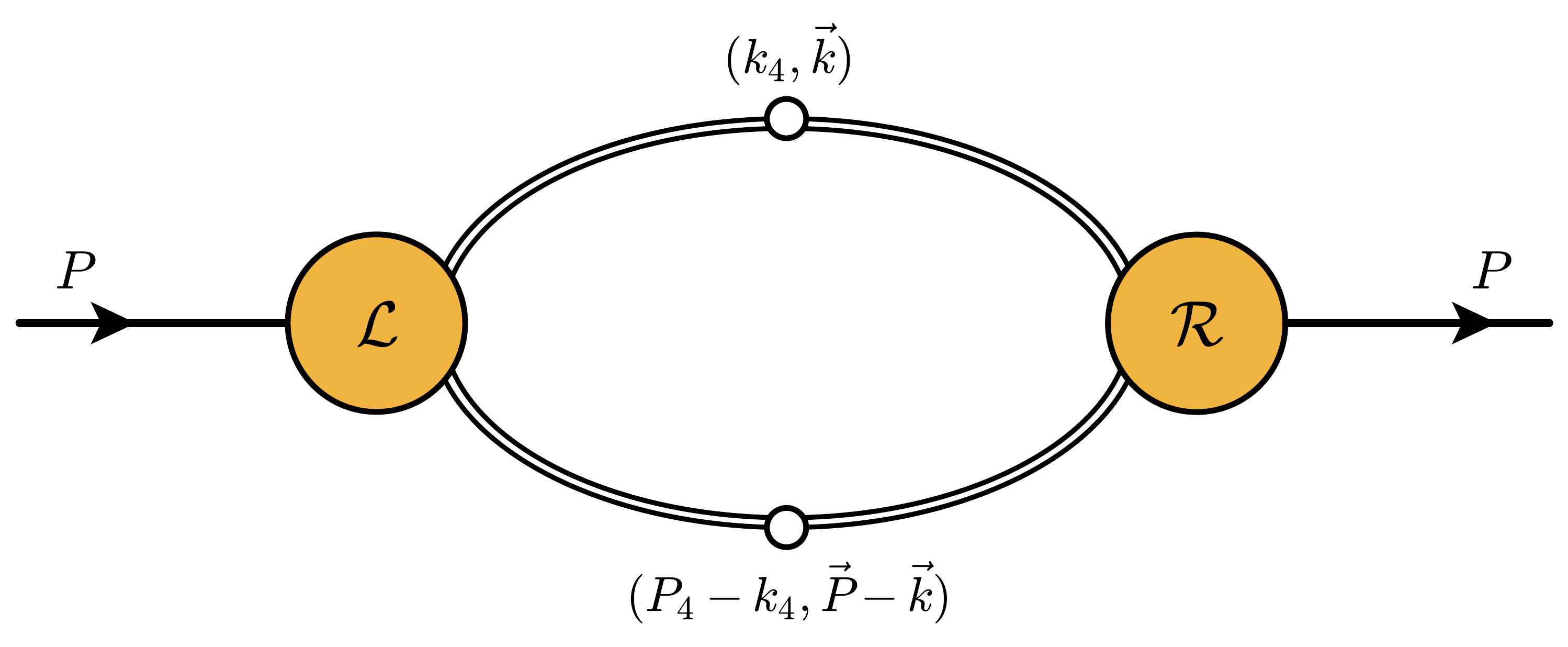}
\caption{The finite-volume two-thermoparticle loop. The double lines represent thermoparticle propagators, the white dots indicate finite-volume momentum and energy sums, and the general functions $\mathcal{L}$ and $\mathcal{R}$ correspond to either endcap functions or Bethe-Salpeter kernels.}
\label{fig:RFT-expansion-loop}
\end{figure}

\noindent
A graphical representation of Eq.~\eqref{eq:L-R-Product_2} is shown in Fig.~\ref{fig:RFT-expansion-loop}. The first step in evaluating $\mathcal{L}\circ \mathcal{R}$ is to compute the Matsubara sum, which is the finite-temperature generalisation of performing the $k_{4}$ integral in the standard approach. A method for performing such sums is to rewrite it in terms of a contour integral where the integer dependence of the function in the sum is taken to be continuous~\cite{Kapusta:2006pm}. The difficultly here is that the singularity structure of the thermoparticle propagator is model-dependent and can potentially be quite complicated, which makes it non-trivial to find a consistent contour that avoids the singularities and their potential branch cuts. \\

\noindent
Instead, one can make use of the representation in Eq.~\eqref{G_TP} and write Eq.~\eqref{eq:L-R-Product_2} in the form      
\begin{align}
\mathcal{L}\circ \mathcal{R} &= \frac{1}{2!} \, \frac{1}{L^3} \!\sum_{\vec{k}\in \frac{2\pi}{L}\,\mathbb{Z}^3} \int_{-\infty}^{\infty} \frac{dq_{0}}{2\pi} \int_{-\infty}^{\infty} \frac{dr_{0}}{2\pi} \, \rho_{\text{TP}}(q_{0},\vec{P}-\vec{k}) \rho_{\text{TP}}(r_{0},\vec{k}) \, \frac{1}{\beta}  \sum_{n \in \mathbb{Z}}  \frac{\mathcal{L}(\frac{2\pi i n}{\beta}, \vec{k})\mathcal{R}(\frac{2\pi i n}{\beta}, \vec{k})}{(iP_{4}-\frac{2\pi i n}{\beta}-q_{0})(\frac{2\pi i n}{\beta}-r_{0})},
\label{eq:L-R-Product_3}
\end{align} 
where the $k_{4}$ singularities of the propagators enter as simple poles at the expense of introducing two additional energy integrals weighted by thermoparticle spectral functions. Now one can evaluate the sum using contour integration by considering the complex function
\begin{align}
F(z)= \pi \cot(\pi z)\frac{\mathcal{L}( \frac{2\pi i z}{\beta},  \vec{k})\mathcal{R}(\frac{2\pi i z}{\beta}, \vec{k})}{(iP_{4}-\frac{2\pi i z}{\beta}-q_{0})(\frac{2\pi i z}{\beta}-r_{0})},
\end{align}
and integrating it along the circular contour $C_{R}$ centred around the origin. The logic here is that the residue of the poles of the $\pi \cot(\pi z)$ factor is precisely the Matsubara sum that one wants to evaluate. The denominator factor has simple poles at $z= -i\beta\frac{iP_{4}-q_{0}}{2\pi}, -i\beta\frac{r_{0}}{2\pi}$, and the functions $\mathcal{L}$, $\mathcal{R}$ will also in general possess singularities. Since $\mathcal{L}$ and $\mathcal{R}$ are either endcap functions or Bethe-Salpeter kernels, they should be bounded for large $|z|$, and hence $F(z)$ must decay like $1/|z|^{2}$ or faster for $|z|\rightarrow \infty$. This implies that the contour integral $\int_{C_{R}}F(z)$ vanishes in the limit $R\rightarrow \infty$, and hence one can use the Cauchy integral theorem to write
\begin{align}
\int_{C_{\infty}}\!  F(z) = 0 = 2\pi i \left[\sum_{n \in \mathbb{Z}} \text{Res}(F(z),z=n) + \sum_{i} \text{Res}(F(z),z=z_{i})  \right] \!, 
\end{align} 
where $z_{i}$ are the remaining singularities. In the vacuum case the singularities from the $\mathcal{L}$ and $\mathcal{R}$ functions are sub-dominant at large volumes. Since this should also hold true at sufficiently small temperatures, in this regime the Matsubara sum can be written entirely in terms of the simple pole residues at $z= -i\beta\frac{iP_{4}-q_{0}}{2\pi}, -i\beta\frac{r_{0}}{2\pi}$, which implies  
\begin{align}
\frac{1}{\beta}  \sum_{n \in \mathbb{Z}}  \frac{\mathcal{L}(\frac{2\pi i n}{\beta}, \vec{k})\mathcal{R}( \frac{2\pi i n}{\beta},  \vec{k})}{2(iP_{4}-\tfrac{2\pi i n}{\beta}-q_{0})(\tfrac{2\pi i n}{\beta}-r_{0})} &= -\frac{\coth\left(\tfrac{\beta q_{0}}{2}\right)\mathcal{L}(iP_{4}-q_{0}, \vec{k})\mathcal{R}(iP_{4}-q_{0}, \vec{k})}{2(iP_{4} - q_{0}- r_{0})} \nonumber \\
& \quad\quad\quad\quad -\frac{\coth\left(\tfrac{\beta r_{0}}{2}\right)\mathcal{L}(r_{0}, \vec{k})\mathcal{R}(r_{0}, \vec{k})   }{2(iP_{4} - q_{0}- r_{0})}.
\label{eq:Matsubara_sum}
\end{align}
The two-thermoparticle loop expression derived so far is restricted to Euclidean energies. To recover the real-time result one must analytically continue: $iP_{4} \rightarrow E + i\epsilon$. Combining Eqs.~\eqref{eq:L-R-Product_3} and~\eqref{eq:Matsubara_sum}, and performing this analytic continuation, one obtains   
\begin{align}
\mathcal{L}\circ \mathcal{R} = &-\frac{1}{4} \, \frac{1}{L^3} \!\sum_{\vec{k}\in \frac{2\pi}{L}\,\mathbb{Z}^3} \int_{-\infty}^{\infty} \frac{dq_{0}}{2\pi} \int_{-\infty}^{\infty} \frac{dr_{0}}{2\pi} \, \mathcal{L}(E-q_{0}, \vec{k}) \frac{\rho_{\text{TP}}(q_{0},\vec{P}-\vec{k}) \rho_{\text{TP}}(r_{0},\vec{k})}{E-q_{0}-r_{0}+i\epsilon}\mathcal{R}(E-q_{0}, \vec{k})\coth\left(\tfrac{\beta q_{0}}{2}\right) \nonumber \\
&-\frac{1}{4} \, \frac{1}{L^3} \!\sum_{\vec{k}\in \frac{2\pi}{L}\,\mathbb{Z}^3} \int_{-\infty}^{\infty} \frac{dq_{0}}{2\pi} \int_{-\infty}^{\infty} \frac{dr_{0}}{2\pi} \, \mathcal{L}(r_{0}, \vec{k}) \frac{\rho_{\text{TP}}(q_{0},\vec{P}-\vec{k}) \rho_{\text{TP}}(r_{0},\vec{k})}{E - q_{0}- r_{0}+i\epsilon}\mathcal{R}(r_{0}, \vec{k}) \coth\left(\tfrac{\beta r_{0}}{2}\right). 
\label{eq:L-R-Product_4}
\end{align}
To classify finite-volume effects one can rewrite the spatial momentum sum as
\begin{align}
\frac{1}{L^3}\!\sum_{\vec{k}\in \frac{2\pi}{L}\,\mathbb{Z}^3}\longrightarrow \int \!  \frac{d^3 \vec{k}}{(2\pi)^3} + \left(\frac{1}{L^3}\!\sum_{\vec{k}\in \frac{2\pi}{L}\,\mathbb{Z}^3}-\int \!  \frac{d^3 \vec{k}}{(2\pi)^3} \right)\!,
\end{align} 
which leads to the decomposition
\begin{align}
\begin{split}
\mathcal{L}\circ \mathcal{R}=(\mathcal{L}\circ \mathcal{R})_{\infty}+(\mathcal{L}\circ \mathcal{R})_{\text{V}},
\label{eq:L-R-finite-infinite-volume-decomp}
\end{split}
\end{align}
where $(\mathcal{L}\circ \mathcal{R})_{\infty}$ is the infinite-volume limit of Eq.~\eqref{eq:L-R-Product_4}, and $(\mathcal{L}\circ \mathcal{R})_{\text{V}}$ now contains all of the finite-volume corrections and has the form 
\begin{align}
(\mathcal{L}\circ \mathcal{R})_{\text{V}} = &-\frac{1}{4} \left(\frac{1}{L^3}\!\sum_{\vec{k}\in \frac{2\pi}{L}\,\mathbb{Z}^3}-\int \!  \frac{d^3 \vec{k}}{(2\pi)^3} \right) \int_{-\infty}^{\infty} \frac{dq_{0}}{2\pi} \int_{-\infty}^{\infty} \frac{dr_{0}}{2\pi} \, \frac{1}{E-q_{0}-r_{0}+i\epsilon} \coth\left(\tfrac{\beta q_{0}}{2}\right)  \nonumber \\
& \quad\quad\quad\quad\quad\quad\quad\quad \times \left[\mathcal{L}(E-q_{0}, \vec{k}) \rho_{\text{TP}}(q_{0},\vec{P}-\vec{k}) \rho_{\text{TP}}(r_{0},\vec{k})\mathcal{R}(E-q_{0}, \vec{k}) \right. \nonumber \\
& \quad\quad\quad\quad\quad\quad\quad\quad\quad\quad\quad\quad\quad\quad + \left. \mathcal{L}(q_{0}, \vec{k}) \rho_{\text{TP}}(q_{0},\vec{k}) \rho_{\text{TP}}(r_{0},\vec{P}-\vec{k}) \mathcal{R}(q_{0}, \vec{k}) \right].
\label{eq:finite-volume-loop-correction}
\end{align}
To simplify this expression further one can perform an analogous partial-wave expansion of the functions $\mathcal{L}$ and $\mathcal{R}$ to that introduced for the scattering amplitude in Sec.~\ref{subsec:partial-wave}, by considering the low-temperature regime where $|\vec{k}|$ is closely clustered around the most probable magnitude of the thermoparticle momentum $|\vec{q}_{\text{TP}}|$. Within this regime the functions have the effective expansions   
\begin{align}
\mathcal{L}(q_0,\vec{k})&=\sqrt{4\pi}\, |\vec{k}|^{\ell'}\, \mathcal{L}_{\ell' m'}(E,\vec{P})\, Y_{\ell' m'}(\hat{k}), \\
\mathcal{R}(q_0,\vec{k})&=\sqrt{4\pi}\, |\vec{k}|^{\ell}\,  Y_{\ell m}^{*}(\hat{k})\,\mathcal{R}_{\ell m}(E,\vec{P}),
\end{align}
where the factor $ |\vec{k}|^{\ell}$ ensures regularity at $\vec{k}=0$ for all $\ell>0$. Using these expansions, Eq.~\eqref{eq:finite-volume-loop-correction} can then be decomposed as
\begin{align}
(\mathcal{L}\circ \mathcal{R})_{\text{V}}= -\mathcal{L}_{\ell' m'}(E,\vec{P})\, F_{\ell' m';\ell m}(E,\vec{P};L,\beta) \, \mathcal{R}_{\ell m}(E,\vec{P}),
\end{align}
where $F_{\ell' m';\ell m}(E,\vec{P};L,\beta)$ is given by
\begin{align}
F_{\ell' m';\ell m}(E,\vec{P};L,\beta) &= \frac{1}{4} \left(\frac{1}{L^3}\!\sum_{\vec{k}\in \frac{2\pi}{L}\,\mathbb{Z}^3}-\int \!  \frac{d^3 \vec{k}}{(2\pi)^3} \right) \int_{-\infty}^{\infty} \frac{dq_{0}}{2\pi} \int_{-\infty}^{\infty} \frac{dr_{0}}{2\pi} \ \frac{4\pi \, Y_{\ell' m'}(\hat{k})\, Y_{\ell m}^{*}(\hat{k})|\vec{k}|^{\ell+\ell'}}{E-q_{0}-r_{0}+i\epsilon}   \nonumber \\
& \quad \times \left[\rho_{\text{TP}}(q_{0},\vec{P}-\vec{k}) \rho_{\text{TP}}(r_{0},\vec{k}) + \rho_{\text{TP}}(q_{0},\vec{k}) \rho_{\text{TP}}(r_{0},\vec{P}-\vec{k})\right]\coth\left(\tfrac{\beta q_{0}}{2}\right).
\label{eq:finite-volume-loop-F-correction}
\end{align}

\noindent
Technically, the continuous $\vec{k}$-integral term within $F_{\ell' m';\ell m}(E,\vec{P};L,\beta)$ in Eq.~\eqref{eq:finite-volume-loop-F-correction} represents a distributional equation in which the energy-dependent component has the decomposition
\begin{align}
\frac{1}{E-q_0-r_0+i\epsilon}=\text{PV}\, \frac{1}{E-q_0-r_0}-i\pi\,\delta(E-q_0-r_0),
\end{align}
where PV is the principal value distribution. With this decomposition one can write 
\begin{align}
F_{\ell' m';\ell m}(E,\vec{P};L,\beta) = F_{\text{PV},\, \ell' m';\ell m}(E,\vec{P};L,\beta)-i F_{\delta,\, \ell' m';\ell m}(E,\vec{P};L,\beta). 
\label{eq:finite-volume-loop-F-correction-PV}
\end{align}
$F_{\text{PV},\, \ell' m';\ell m}(E,\vec{P};L,\beta)$ contains the real part of both the sum and integral components in Eq.~\eqref{eq:finite-volume-loop-F-correction}, whereas $F_{\delta,\, \ell' m';\ell m}(E,\vec{P};L,\beta)$ involves only the imaginary part of the $\vec{k}$-integral, and is equal to $(1-e^{-\beta E})\varrho_{\ell m;\ell' m', \text{TP}}(E,\vec{P})$, where $\varrho_{\ell m;\ell' m', \text{TP}}(E,\vec{P})$ is the physical two-thermoparticle phase space defined in Eq.~\eqref{eq:phase-space-th-general-P}. In the standard formalism energy positivity ($E>0$) is implicitly assumed, and hence this proportionality factor approaches one in the zero-temperature $\beta \rightarrow \infty$ limit. Taking into account the threshold properties $q_{0}^{2} \geq m^{2}$, $r_{0}^{2} \geq m^{2}$ of the thermoparticle spectral functions within the integration regions, $F_{\text{PV},\, \ell' m';\ell m}(E,\vec{P};L,\beta)$ can be written in the form 
\begin{align}
F_{\text{PV},\, \ell' m';\ell m}(E,\vec{P};L,\beta) &= \pi \left(\frac{1}{L^3}\!\sum_{\vec{k}\in \frac{2\pi}{L}\,\mathbb{Z}^3}-\int \!  \frac{d^3 \vec{k}}{(2\pi)^3} \right) \int_{m}^{\infty} \frac{dq_{0}}{2\pi} \ \text{PV} \!\int_{m}^{\infty} \frac{dr_{0}}{2\pi} \,  Y_{\ell' m'}(\hat{k})\, Y_{\ell m}^{*}(\hat{k})|\vec{k}|^{\ell+\ell'}   \nonumber \\
& \quad \times \coth\left(\tfrac{\beta q_{0}}{2}\right) \left[ \frac{\rho_{\text{TP}}(q_{0},\vec{P}-\vec{k}) \rho_{\text{TP}}(r_{0},\vec{k}) + \rho_{\text{TP}}(q_{0},\vec{k}) \rho_{\text{TP}}(r_{0},\vec{P}-\vec{k})}{E-q_{0}-r_{0}}  \right.  \nonumber \\
&\quad\quad\quad\quad - \frac{\rho_{\text{TP}}(q_{0},\vec{P}-\vec{k}) \rho_{\text{TP}}(r_{0},\vec{k}) + \rho_{\text{TP}}(q_{0},\vec{k}) \rho_{\text{TP}}(r_{0},\vec{P}-\vec{k})}{E-q_{0}+r_{0}} \nonumber \\
&\quad\quad\quad\quad + \frac{\rho_{\text{TP}}(q_{0},\vec{P}-\vec{k}) \rho_{\text{TP}}(r_{0},\vec{k}) + \rho_{\text{TP}}(q_{0},\vec{k}) \rho_{\text{TP}}(r_{0},\vec{P}-\vec{k})}{E+q_{0}-r_{0}} \nonumber \\
&\left. \quad\quad\quad\quad - \frac{\rho_{\text{TP}}(q_{0},\vec{P}-\vec{k}) \rho_{\text{TP}}(r_{0},\vec{k}) + \rho_{\text{TP}}(q_{0},\vec{k}) \rho_{\text{TP}}(r_{0},\vec{P}-\vec{k})}{E+q_{0}+r_{0}} \right].
\label{eq:finite-volume-loop-F-correction-PV-3} 
\end{align}
Equation~\eqref{eq:finite-volume-loop-F-correction-PV-3} contains all of the leading finite-volume effects of the two-thermoparticle loop for general total momentum $\vec{P}$, and corresponds to the finite-temperature generalisation of the real $F$ function derived in Ref.~\cite{Hansen:2012tf}. In the zero-temperature limit: $\rho_{\text{TP}}(q_{0},\vec{k}) \rightarrow 2\pi \epsilon(q_{0})\delta(q_{0}^{2}-\omega_{k}^{2})$, and hence Eq.~\eqref{eq:finite-volume-loop-F-correction-PV-3} reduces to 
\begin{align}
F_{\ell' m';\ell m}(E,\vec{P};L,\beta) &\xrightarrow{\beta\rightarrow \infty}{} \left(\frac{1}{L^3}\!\sum_{\vec{k}\in \frac{2\pi}{L}\,\mathbb{Z}^3}-\int \!  \frac{d^3 \vec{k}}{(2\pi)^3} \right) \left[\frac{2\pi \, Y_{\ell' m'}(\hat{k}^{*})\, Y_{\ell m}^{*}(\hat{k}^{*})|\vec{k}|^{\ell+\ell'}}{2\omega_{k}2\omega_{Pk}\, (E-\omega_{k}-\omega_{Pk})}  \right.  \nonumber \\
&\quad\quad\quad\quad\quad\quad\quad\quad\quad \left.  - \frac{2\pi \, Y_{\ell' m'}(\hat{k}^{*})\, Y_{\ell m}^{*}(\hat{k}^{*})|\vec{k}|^{\ell+\ell'}}{2\omega_{k}2\omega_{Pk}\, (E+\omega_{k}+\omega_{Pk})} \right],
\label{eq:finite-volume-loop-F-vac}
\end{align}
where $\omega_{k}= \sqrt{|\vec{k}|^{2}+m^{2}}$,  $\omega_{Pk}= \sqrt{|\vec{P}-\vec{k}|^{2}+m^{2}}$, and $\hat{k}^{*}$ is the angular direction in the two-particle centre-of-momentum frame. The first term in Eq.~\eqref{eq:finite-volume-loop-F-vac} coincides with the standard result for the real part of the zero-temperature $F$ function~\cite{Kim:2005gf,Hansen:2012tf}. Since the second term does not have poles for $E \geq 0$, it leads to exponentially suppressed finite-volume corrections, and is therefore neglected in the standard approach. \\

\noindent
The final components to consider in the skeleton expansion of $C_{L,\beta}(P_4,\vec{P})$ are the finite-volume thermal Bethe-Salpeter kernels $B_{L,\beta}(k'_1,k'_2;k_1,k_2)$. In the zero-temperature case the Bethe-Salpeter kernels which contain more than two internal particle lines cannot have all of these lines simultaneously on shell for energies $E<3m$, and hence they do not generate poles. As a result, for $E<3m$ these kernels lead to exponentially-suppressed finite-volume corrections\footnote{Technically, this is true up to a caveat related to the left-hand cut due to $t$ and $u$-channel exchange addressed in Ref.~\cite{Raposo:2023oru}. This was not considered in Refs.~\cite{Kim:2005gf,Hansen:2012tf}, and will also not be taken into account in this analysis.}. As discussed previously, the singularity structure of finite-temperature propagators is more general, and may no longer contain simple poles. This means that the separation between power-law-like and exponential-like finite-volume effects can potentially be lost. Nevertheless, at sufficiently low temperatures, i.e. large but finite $L_{\tau}$, the thermoparticle spectral function becomes increasingly peaked around the vacuum singularity, and so we expect that the finite-volume effects from contributions of more than two thermoparticle propagators in the Bethe-Salpeter kernel will continue to be sub-leading compared to those from the two-thermoparticle loop when $E<3m$. Since only leading finite-volume effects are considered, this means that one can replace all finite-volume kernels $B_{L,\beta}(k'_1,k'_2;k_1,k_2)$ appearing in the skeleton expansion of $C_{L,\beta}(P_4,\vec{P})$ by their infinite-volume limits $B_{\infty,\beta}(k'_1,k'_2;k_1,k_2)$.

\subsection{Finite-temperature two-particle quantisation condition}

Using the decomposition in Eq.~\eqref{eq:L-R-finite-infinite-volume-decomp} one can reorder all the terms inside the expansion in Eq.~\eqref{eq:skeleton-expansion-finite-kernels} with respect to the number of $F(E,\vec{P};L,\beta)$ insertions. After performing a partial-wave expansion of $F$ and the Bethe-Salpeter kernels, the resulting reordered expansion can be written 
\begin{align}
C_{L,\beta}(E,\vec{P})=C_{\infty,\beta}(E,\vec{P})-\mathcal{A}(E,\vec{P})\, F(E,\vec{P};L,\beta)\, \sum_{n=0}^{\infty}\, \left[-\mathcal{M}_{\text{TP}}(E,\vec{P})\,F(E,\vec{P};L,\beta)\right]^{n} \, \mathcal{A}^{\dagger}(E,\vec{P}),
\label{eq:skeleton-partial-wave}
\end{align}
where $\mathcal{M}_{\text{TP}}(E,\vec{P})$ is the infinite-volume two-thermoparticle scattering amplitude defined at arbitrary values of $(E,\vec{P})$, and all partial wave indices are kept implicit. The dressed endcap functions $\mathcal{A}(E,\vec{P})$, $\mathcal{A}^{\dagger}(E,\vec{P})$ and the scattering amplitude $\mathcal{M}_{\text{TP}}(E,\vec{P})$ have the form
\begin{align}
\mathcal{A}(E,\vec{P})&=\tilde{\sigma}(E,\vec{P}) \sum_{n=0}^{\infty}\, \left[-F(E,\vec{P};\infty,\beta)\, iB(E,\vec{P};\infty,\beta)\right]^{n}, \\
\mathcal{A}^{\dagger}(E,\vec{P})&=\sum_{n=0}^{\infty}\, \left[iB(E,\vec{P};\infty,\beta)\, F(E,\vec{P};\infty,\beta)\right]^{n} \tilde{\sigma}^{\dagger}(E,\vec{P}), \\
\mathcal{M}_{\text{TP}}(E,\vec{P})&=iB(E,\vec{P};\infty,\beta) \sum_{n=0}^{\infty}\, \left[-F(E,\vec{P};\infty,\beta)\, iB(E,\vec{P};\infty,\beta)\right]^{n} \!.
\end{align}
Computing the geometric series in Eq.~\eqref{eq:skeleton-partial-wave} one then obtains
\begin{align}
C_{L,\beta}(E,\vec{P})=C_{\infty,\beta}(E,\vec{P})-\mathcal{A}(E,\vec{P})\, \frac{1}{F^{-1}(E,\vec{P};L,\beta)+\mathcal{M}_{\text{TP}}(E,\vec{P})}\, \mathcal{A}^{\dagger}(E,\vec{P}).
\label{eq:skeleton-geometric-series}
\end{align}
Since $C_{L,\beta}(E,\vec{P})$ has poles at the finite-volume thermal energy levels $\mathcal{E}(\vec{P};L,\beta)$, whereas $C_{\infty,\beta}(E,\vec{P})$ does not, these pole singularities can only arise from the second term in Eq.~\eqref{eq:skeleton-geometric-series}. This implies that $[F^{-1}(E,\vec{P};L,\beta)+\mathcal{M}_{\text{TP}}(E,\vec{P})]$ must be a singular matrix in partial-wave space for $E=\mathcal{E}(\vec{P};L,\beta)$, and hence satisfy the condition   
\begin{align}
\text{det}\,[F^{-1}(\mathcal{E},\vec{P};L,\beta) + \mathcal{M}_{\text{TP}}(\mathcal{E},\vec{P})]=0.
\label{eq:qc2-1}
\end{align}
Alternatively, Eq.~\eqref{eq:qc2-1} can be expressed in terms of purely real quantities as
\begin{align}
\text{det} \, [F_{\text{PV}}^{-1}(\mathcal{E},\vec{P};L,\beta) + \mathcal{K}_{\text{TP}}(\mathcal{E},\vec{P})]=0,
\label{eq:qc2-2}
\end{align}
where $\mathcal{K}_{\text{TP}}$ is the analogue of the infinite-volume two-thermoparticle $K$-matrix in Eq.~\eqref{eq:M_K}, but defined for arbitrary $(E,\vec{P})$, and $F_{\text{PV}}$ is given in Eq.~\eqref{eq:finite-volume-loop-F-correction-PV-3}. Equation~\eqref{eq:qc2-1} represents the finite-temperature generalisation of the standard vacuum two-particle quantisation condition defined in Eq.~\eqref{eq:vacuum-quantisation-condition}. As in the vacuum case~\cite{Kim:2005gf,Hansen:2012tf}, Eq.~\eqref{eq:qc2-2} can be solved at each energy $\mathcal{E}$ in order to constrain the form of $\mathcal{K}_{\text{TP}}(\mathcal{E},\vec{P})$. However, the temperature dependence introduces the additional subtlety that Eq.~\eqref{eq:qc2-2} must now be solved \textit{separately} for each $\vec{P}$, due to the loss of boost invariance. This results in different constraints on the $K$-matrix for each momentum frame. \\

\noindent
The quantisation condition in Eq.~\eqref{eq:qc2-2} is defined in the partial-wave basis, although this basis only truly exists for infinite spatial volumes, where rotational symmetry is exact. In a finite cubic box this symmetry is broken, and correlation functions instead belong to irreducible representations (irreps) $\Gamma\in \{A_1^+,\, T_1^+,\,...\}$ of the cubic group rather than to definite partial waves. To project Eq.~\eqref{eq:qc2-2} to these irreps one can apply the same subduction procedure proposed in Ref.~\cite{Dudek:2010wm} for the vacuum case, since these irreps are independent of the temporal lattice extent. Performing this projection gives    
\begin{align}
&\text{det} \, [(F_{\text{PV}}^{(\Gamma,\, \mu)})^{-1}(\mathcal{E},\vec{P};L,\beta) + \mathcal{K}_{\text{TP}}(\mathcal{E},\vec{P})]=0,  
\label{eq:qc2-irrep}  \\
&F_{\text{PV},\, \ell' \ell}^{(\Gamma,\, \mu)}(\mathcal{E},\vec{P};L,\beta)=\sum_{m,m'}\, S^{(\Gamma,\, \mu)}_{\ell' m'}\, F_{\text{PV},\, \ell' m';\ell m}(\mathcal{E},\vec{P};L,\beta)\,  S^{(\Gamma,\, \mu)}_{\ell m},
\label{eq:subduct}
\end{align}
where $\mu\in \{1,...,\text{dim}(\Gamma)\}$ labels the rows of the specified irrep $\Gamma$, and $S^{(\Gamma,\, \mu)}_{\ell' m'}$ are the corresponding subduction coefficients. An important characteristic of cubic irreps is that they contain an infinite number of allowed orbital angular momenta $\ell$, and hence $F_{\text{PV},\, \ell' \ell}^{(\Gamma,\, \mu)}(\mathcal{E},\vec{P};L,\beta)$ are infinite-dimensional matrices in this representation space. For practical applications one must therefore truncate these matrices, which is usually done by considering only the lowest-lying partial waves in any given irrep.

\subsection{Analysis procedure for finite-temperature resonances} 
\label{subsec:algorithm}

The results derived in this section emphasise that the finite-temperature generalisation of the two-particle quantisation condition in Eq.~\eqref{eq:qc2-1} is strongly model dependent. To compute $F(\mathcal{E},\vec{P};L,\beta)$ one needs to establish the properties of the thermoparticle excitations, in particular the form of their damping factors $D_{m,\beta}(\vec{x})$. Given $D_{m,\beta}(\vec{x})$, this determines the spectral function $\rho_{\text{PT}}(\omega,\vec{p})$, and hence $F(\mathcal{E},\vec{P};L,\beta)$. As discussed in Secs.~\ref{sec:TP_basics} and~\ref{sec:infinite-volume-scattering}, in any given theory the form of $D_{m,\beta}(\vec{x})$ can be uniquely fixed by the dynamical equations. In practice though, it often more convenient to extract the form of $D_{m,\beta}(\vec{x})$ directly from correlation function data~\cite{Lowdon:2022keu,Lowdon:2022xcl,Bala:2023iqu,Lowdon:2024atn,Ali:2026ehk,Lowdon:2025fyb,Lowdon:2025ait}. In lattice QCD this involves the analysis of two-point spatial and temporal correlators of single-hadron interpolating operators~\cite{Lowdon:2022xcl,Bala:2023iqu}. With this in mind, this leads to the following procedure for computing the properties of resonances at finite temperature: 

\begin{enumerate}

\item Compute the spatial $C(z)$ and temporal $C(\tau)$ correlation functions of single-particle interpolating operators, and extract the damping factor $D_{m,\beta}(\vec{x})$ of the lowest-lying state (of vacuum mass $m$) at different temperatures by varying the temporal lattice size.  

\item Determine the energy levels $\mathcal{E}(\vec{P};L,\beta)$ from the temporal correlation functions of the two-particle interpolating operators for different frames and irreps at these temperatures.   

\item Solve the quantisation condition in Eq.~\eqref{eq:qc2-irrep} for fixed $\vec{P}$ with a chosen parametrisation of $\mathcal{K}_{\text{TP}}(\mathcal{E},\vec{P})$, and fit to the energy levels computed in the previous step.

\item Compute the phase space $\varrho_{\text{TP}}(E,\vec{P})$ via Eq.~\eqref{eq:phase-space-th-general-P}, and use Eq.~\eqref{eq:M_K} together with $\mathcal{K}_{\text{TP}}(\mathcal{E},\vec{P})$ from the previous step to determine the scattering amplitude $\mathcal{M}_{\text{TP}}(E,\vec{P})$.

\item Extract the resonance parameters and their temperature dependence from the singularity structure of $\mathcal{M}_{\text{TP}}(E,\vec{P})$.

\end{enumerate}

For consistency, these steps must be carried out on the same lattice ensembles in order to preserve the statistical correlations. With the simplest choice of parametrisation the final outcome of the procedure will be the temperature-dependent resonance mass $M_{p}(\beta)$ and width $\Gamma(\beta)$, which will approach the values obtained in the standard procedure for sufficiently large temporal lattice sizes, i.e. $\beta = L_{\tau} \rightarrow \infty$.

\subsection{Two-thermoparticle kinematic function in the rest frame} 

To illustrate the differences between the real part of the kinematic function at finite-temperature and in vacuum we consider the rest frame $\vec{P}=0$ and $D_{m,\beta}(\vec{x})= e^{-\gamma |\vec{x}|}$ with $\gamma=T$, as in Sec.~\ref{subsec:rest-frame-phase-space}. Furthermore, for simplicity we also fix $\Gamma$ to the $A_1^{+}$ irrep and consider only the s-wave components in the partial-wave expansion, i.e. contributions with $\ell,\ell'=0$. Using the symmetry of $F_{\text{PV}}$ under the interchange $E \rightarrow -E$, and taking into account the location of the poles in the integrand denominators, it follows from Eqs.~\eqref{eq:finite-volume-loop-F-correction-PV-3} and~\eqref{eq:subduct} that $F_{\text{PV},\, 00}^{A_1^+}(E) := F_{\text{PV},\, 00}^{A_1^+}(E,\vec{P}=0;L,\beta)$ can be written in the form
\begin{align}
&F_{\text{PV},\, 00}^{A_1^+}(E)= -\frac{1}{2}  \left(\frac{1}{L^3}\!\sum_{\vec{k}\in \frac{2\pi}{L}\,\mathbb{Z}^3}-\int \!  \frac{d^3 \vec{k}}{(2\pi)^3} \right) \left[ \int_{m}^{\infty} \frac{dq_{0}}{2\pi} \int_{m}^{\infty}  \frac{dr_{0}}{2\pi} \, \rho_{\text{TP}}(q_{0},\vec{k}) \rho_{\text{TP}}(r_{0},\vec{k})  \frac{\coth\left(\frac{q_{0}}{2T}\right)}{|E| + q_{0}+ r_{0}} \right. \nonumber \\
&+  \left( \int_{m}^{|E|+m} \frac{dq_{0}}{2\pi} \int_{m}^{\infty}  \frac{dr_{0}}{2\pi} + \int_{|E|+m}^{\infty} \frac{dq_{0}}{2\pi} \ \text{PV} \!\int_{m}^{\infty}\frac{dr_{0}}{2\pi} \right) \, \rho_{\text{TP}}(q_{0},\vec{k}) \rho_{\text{TP}}(r_{0},\vec{k})  \frac{ \coth\left(\frac{q_{0}}{2T}\right)}{|E| - q_{0}+ r_{0}} \nonumber \\
&- \int_{m}^{\infty} \frac{dq_{0}}{2\pi} \ \text{PV} \! \int_{m}^{\infty}\frac{dr_{0}}{2\pi} \, \rho_{\text{TP}}(q_{0},\vec{k}) \rho_{\text{TP}}(r_{0},\vec{k})   \frac{\coth\left(\frac{q_{0}}{2T}\right)}{|E| + q_{0}- r_{0}} \nonumber \\
&+ \theta(|E|-2m) \left( \int_{|E|-m}^{\infty} \frac{dq_{0}}{2\pi} \int_{m}^{\infty}  \frac{dr_{0}}{2\pi} + \int_{m}^{|E|-m} \frac{dq_{0}}{2\pi} \ \text{PV} \!\int_{m}^{\infty}\frac{dr_{0}}{2\pi}\right) \, \rho_{\text{TP}}(q_{0},\vec{k}) \rho_{\text{TP}}(r_{0},\vec{k})  \frac{\coth\left(\frac{q_{0}}{2T}\right)}{q_{0}+r_{0}-|E|} \nonumber \\
&+ \left. \theta(2m-|E|) \int_{m}^{\infty} \frac{dq_{0}}{2\pi} \int_{m}^{\infty}  \frac{dr_{0}}{2\pi} \, \rho_{\text{TP}}(q_{0},\vec{k}) \rho_{\text{TP}}(r_{0},\vec{k})  \frac{\coth\left(\frac{q_{0}}{2T}\right)}{q_{0}+ r_{0}-|E|} \right].
\label{eq:finite-volume-loop-principal-rest-frame}
\end{align}
 
\begin{figure}[t!]
\centering
\includegraphics[width=0.49\textwidth]{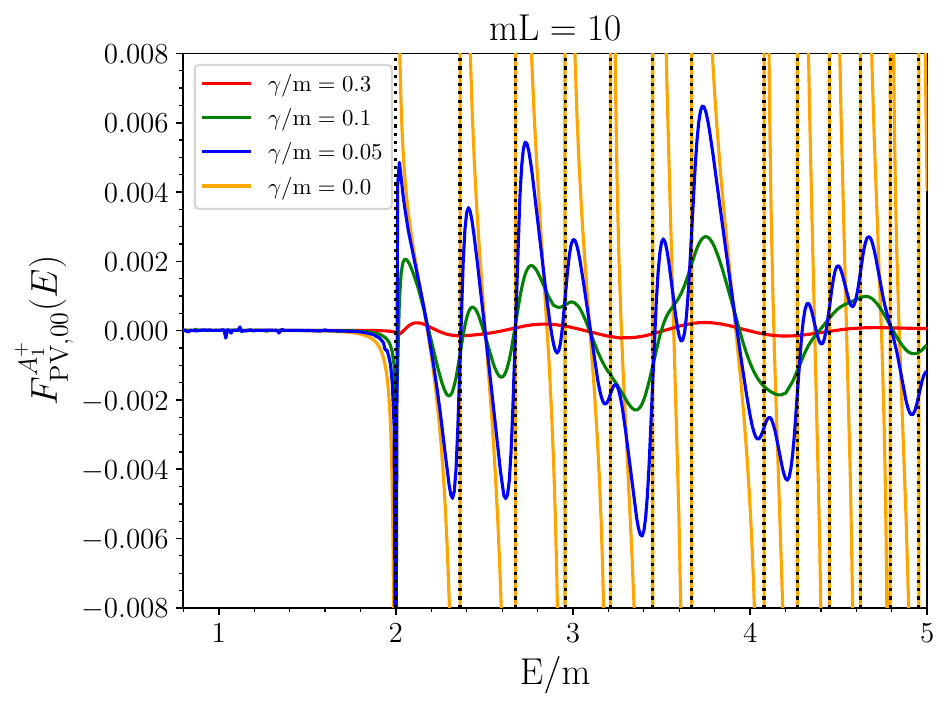}
\includegraphics[width=0.49\textwidth]{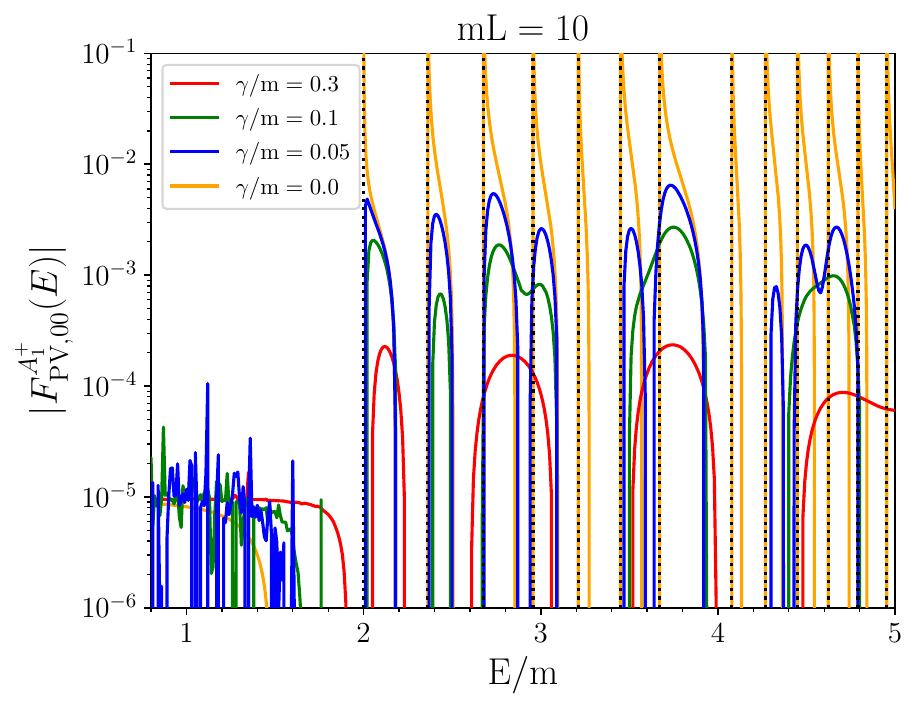}
\caption{Real part of the rest-frame kinematic function $F_{\text{PV},\, 00}^{A_1^+}(E)$ projected to the $A_1^{+}$ cubic volume irrep, with an exponential damping factor $D_{m,\beta}(\vec{x}) =  e^{-\gamma |\vec{x}|}$ at $\gamma/m=$ 0, 0.05, 0.1, 0.3, and $mL=10$.}
\label{fig:Lüscher-L-10}
\end{figure}

\begin{figure}[t!]
\centering
\includegraphics[width=0.49\textwidth]{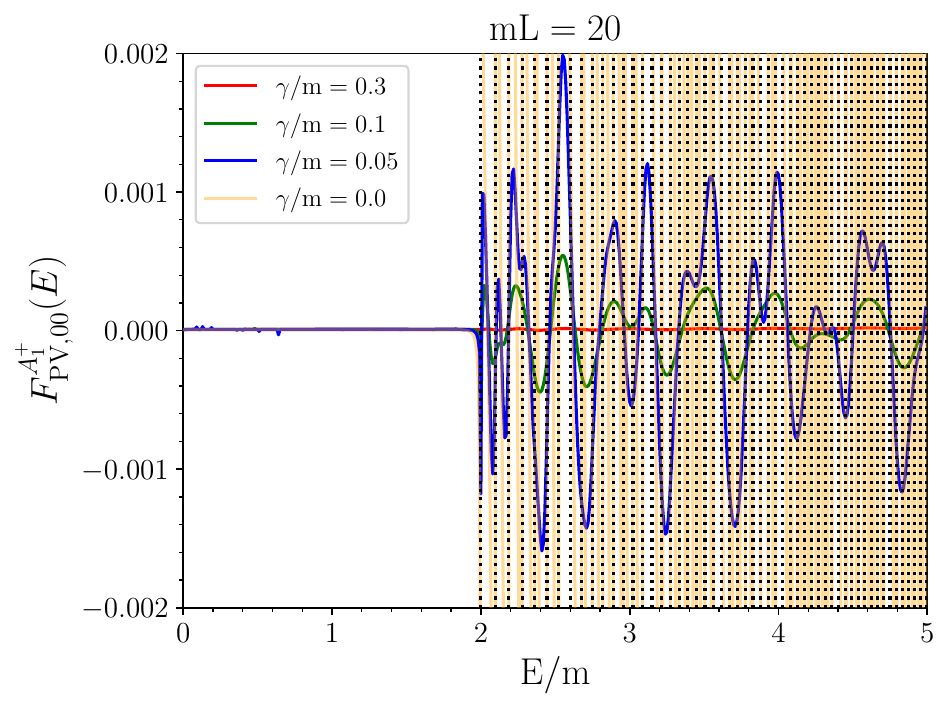} 
\includegraphics[width=0.49\textwidth]{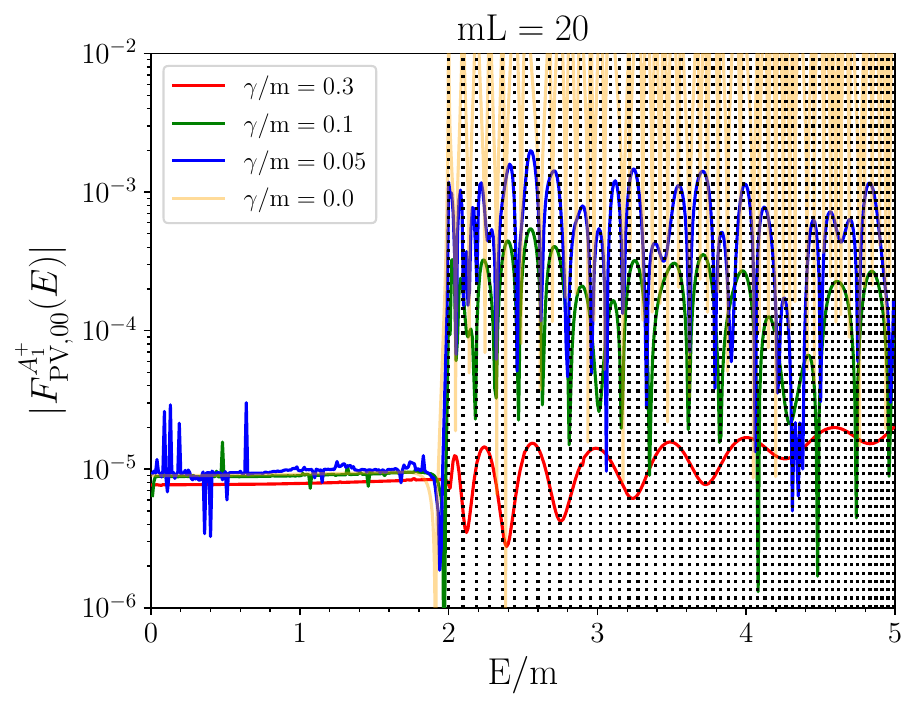}
\caption{Real part of the rest-frame kinematic function $F_{\text{PV},\, 00}^{A_1^+}(E)$ projected to the $A_1^{+}$ cubic volume irrep, with an exponential damping factor $D_{m,\beta}(\vec{x}) =  e^{-\gamma |\vec{x}|}$ at $\gamma/m=$ 0, 0.05, 0.1, 0.3, and $mL=20$.}
\label{fig:Lüscher-L-20}
\end{figure}

In order to evaluate Eq.~\eqref{eq:finite-volume-loop-principal-rest-frame} numerically we computed each of the integrals separately using the multi-dimensional integration package \textit{cubature}~\cite{cubature_c}. We imposed finite cutoffs $(\Lambda_k, \Lambda_{q_0}, \Lambda_{r_0})$ on the sum and integral over $\vec{k}$, as well as the integrals over $q_0$ and $r_0$, and the principal value integrals were evaluated by introducing a small regulator $\varepsilon=10^{-6}$, as described in Appendix~\ref{sec:Appendix}. The cutoffs and number of integration steps were chosen so as to avoid large numerical fluctuations, particularly in the energy region of interest $E\geq m$. A more detailed account of these selection procedures can be found in Appendix~\ref{sec:Appendix}. In Figs.~\ref{fig:Lüscher-L-5}-\ref{fig:Lüscher-L-20} $F_{\text{PV},\, 00}^{A_1^+}(E)$ is plotted over the parameter range $\gamma/m\in \{0.0,0.05,0.1,0.3\}$ for three different box sizes $mL=$ 5, 10, 20. In the vacuum case $\gamma/m=0$, which coincides with the standard two-particle quantisation formalism, one observes poles for $E\geq 2m$. These poles correspond to the energy levels of non-interacting two-particle states, and are displayed as vertical dotted black lines in the plots. Below threshold ($E<2m$) the function is non-vanishing for $\gamma/m=0$. This non-physical behaviour reflects the fact that closed decay channels can contribute below threshold because of finite-volume effects~\cite{Briceno:2017max}. Nevertheless, this effect is exponentially suppressed with increasing volume, as seen in Figs.~\ref{fig:Lüscher-L-5}-\ref{fig:Lüscher-L-20}. \\

\begin{figure}[t!]
\centering
\includegraphics[width=0.49\textwidth]{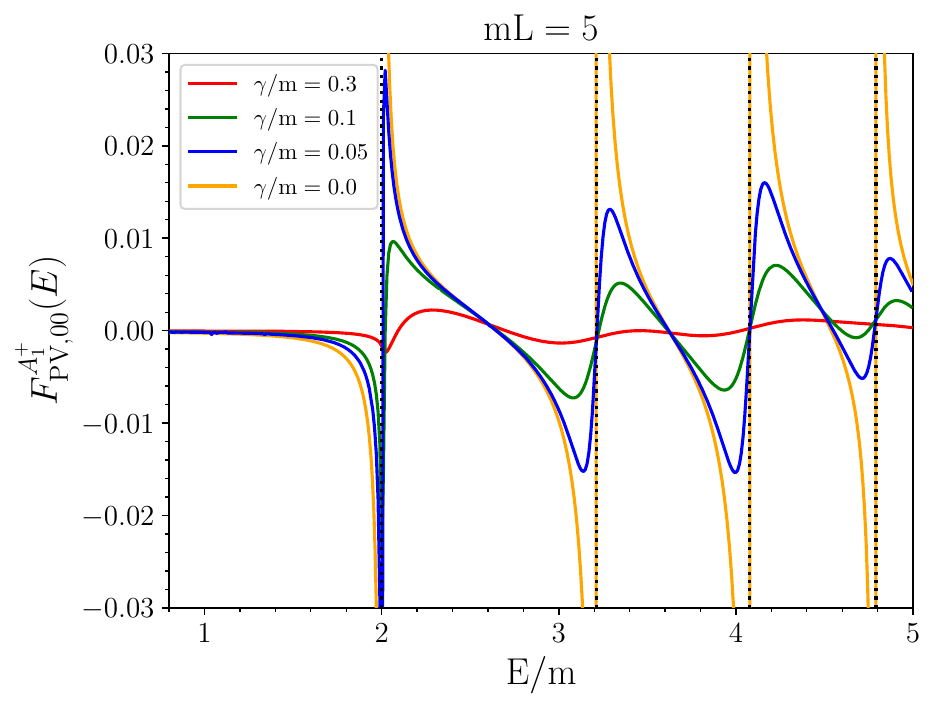}
\includegraphics[width=0.49\textwidth]{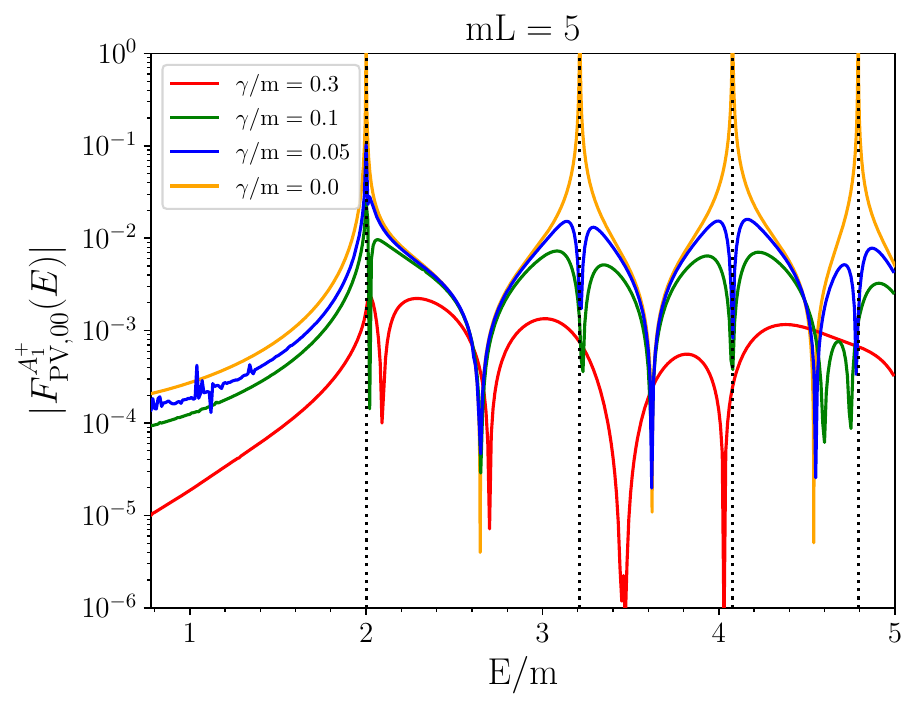}
\caption{Real part of the rest-frame kinematic function $F_{\text{PV},\, 00}^{A_1^+}(E)$ projected to the $A_1^{+}$ cubic volume irrep, with an exponential damping factor $D_{m,\beta}(\vec{x}) =  e^{-\gamma |\vec{x}|}$ at $\gamma/m=$ 0, 0.05, 0.1, 0.3, and $mL=5$.}
\label{fig:Lüscher-L-5}
\end{figure}

\begin{figure}[t!]
\centering
\includegraphics[width=0.5\textwidth]{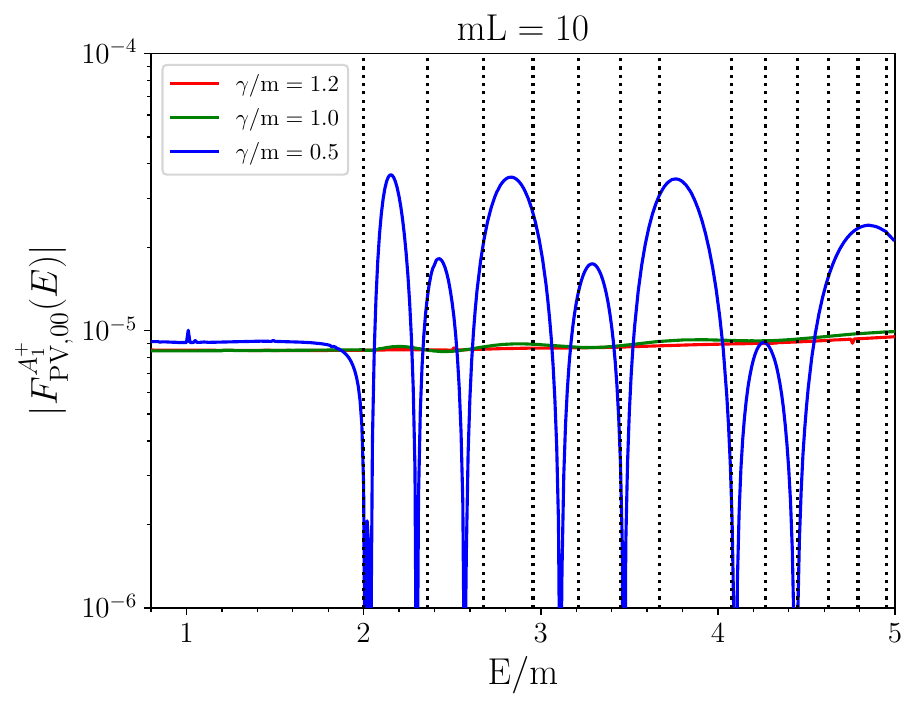}
\caption{Real part of the rest-frame kinematic function $F_{\text{PV},\, 00}^{A_1^+}(E)$ projected to the $A_1^{+}$ cubic volume irrep, with an exponential damping factor $D_{m,\beta}(\vec{x}) =  e^{-\gamma |\vec{x}|}$ at $\gamma/m=$ 0.5, 1.0, 1.2, and $mL=10$.}
\label{fig:Lüscher-high}
\end{figure}

\noindent
As soon as the temperature is non-vanishing, i.e. $\gamma/m>0$, the two-particle poles for $E\geq 2m$ turn into pairs of finite peaks above and below the vacuum pole energies. The distance between these peaks increases with temperature, as can be best seen in the right-panels of Figs.~\ref{fig:Lüscher-L-5}-\ref{fig:Lüscher-L-20}, where $|F_{\text{PV},\, 00}^{A_1^+}(E)|$ is plotted on a logarithmic scale. The peak amplitudes also decrease with increasing $\gamma/m$, and for $\gamma/m>0.5$ these peak structures become increasingly less resolvable, as shown in Fig.~\ref{fig:Lüscher-high}. In the below-threshold region there is an enhanced suppression relative to the $\gamma/m=0$ case, and this gets more pronounced for increasing $\gamma/m$. These features are universal across all of the three box sizes $mL=5,10,20$, the only difference is the density of the peaks, which increases as a function of $mL$ because of the larger number of energy levels. The observed characteristics for $\gamma/m>0$ arise from the fact that the thermoparticle spectral function $\rho_{\text{TP}}(\omega,\vec{p})$ has a broadened peak-like structure around the vacuum singularity $\omega = \sqrt{|\vec{p}|^{2}+m^{2}}$, and hence the poles of the thermoparticle propagators are off the real axis. The real momentum sums and integrals can therefore be performed without encountering these poles, and hence $F_{\text{PV},\, 00}^{A_1^+}(E)$ no longer possesses divergences in $E$. As the temperature increases the thermoparticle spectral function peaks become broadened and flattened, which captures the increased interactions that these states experience with the surrounding thermal medium. These collisional broadening effects explain the damping of the peaks in Figs.~\ref{fig:Lüscher-L-5}-\ref{fig:Lüscher-high} for $E \geq 2m$, as well as the enhanced below-threshold suppression. The small fluctuations seen in the logarithmic-scale plots of $|F_{\text{PV},\, 00}^{A_1^+}(E)|$, particularly for $E < 2m$, are numerical integration artefacts, as discussed in Appendix~\ref{sec:Appendix}. When the temperature is sufficiently large the spectral peaks are completely screened, which is reflected by the absence of any structure in Fig.~\ref{fig:Lüscher-high} for $\gamma/m \gtrsim 1$. 

\subsection{Finite temporal size effects}

The fact that the temporal size $L_{\tau}$ is finite in any practical lattice simulation implies that the corresponding system is always subject to a non-vanishing temperature $T=1/L_{\tau}$. In contrast to finite spatial volume $L>0$, the extent to which finite-$L_{\tau}$ effects manifest themselves depends on the specific dynamics of the system. This is because temperature is intrinsically connected to the spectrum of the theory via the KMS condition, which for finite volumes implies the thermal correlator decomposition in Eq.~\eqref{KMS_corr}. Therefore, in order to analyse either dynamical effects in thermal systems, or finite-$L_{\tau}$ corrections to vacuum lattice observables, this necessarily requires one to understand the finite-temperature dynamics of the system. In the quantisation condition in Eq.~\eqref{eq:qc2-1} this dynamical dependence is reflected in the fact that both the kinematic $F$ function \textit{and} the scattering amplitude depend on the properties of the thermoparticle scattering states. In the $L_{\tau}\rightarrow \infty$ limit one recovers the standard condition in Eq.~\eqref{eq:vacuum-quantisation-condition}, where the $F$ function is model independent. Although one expects Eq.~\eqref{eq:vacuum-quantisation-condition} should hold for sufficiently large $L_{\tau}$, Eq.~\eqref{eq:qc2-1} can be used to assess the impact of finite-$L_{\tau}$ corrections. For example, in Figs.~\ref{fig:Lüscher-L-5}-\ref{fig:Lüscher-high} one can see that $F_{\text{PV}}(E,\vec{P};L,\beta)$ no longer has poles at the non-interacting particle energies, but instead has finite peaks whose positions and widths depend on $\beta=L_{\tau}$. These effects could potentially lead to non-negligible differences even for large $L_{\tau}$, especially for systems where the thermal energy levels lie close to the non-interacting two-particle levels. An investigation of these effects will be the subject of a future work.

\section{Conclusions} 
\label{sec:Conclusion}

In this work we used the concept of thermoparticles in order to generalise the notion of scattering to finite temperature. Firstly, we derived a thermal optical theorem for the scattering amplitude of thermoparticle states, and established that the associated two-thermoparticle phase space $\varrho_{\text{TP}}(E)$ is non-vanishing both below the vacuum two-particle threshold $E=2m$, and for negative energies $E<0$. These characteristics reflect the fact that there is always the potential for either low-energy thermal excitations, or to create hole-like states by extracting energy from the medium itself, although both of these are thermodynamically suppressed. Analytically, $\varrho_{\text{TP}}(E)$ decomposes into above-threshold $E>2m$ and purely thermal non-threshold components. At higher temperatures, the non-threshold components become increasingly enhanced, and at some point the signature of the vacuum threshold is effectively screened. Physically, this corresponds to the regime in which thermoparticles are overwhelmed by collective thermal excitations, and therefore no longer represent the dominant degrees of freedom in the system. \\ 

\noindent
We subsequently derived a finite-temperature generalisation of the finite-volume two-particle quantisation condition, which relates the thermal scattering amplitude to the energy levels computed on lattices with finite spatial $L$ and temporal extent $L_{\tau}$. All leading-order finite-volume effects are contained within the kinematic function $F(E,\vec{P};L,\beta)$, which no longer has poles at the non-interacting particle energies, as in the vacuum $\beta=L_{\tau}=\infty$ case, but instead has finite peaks. This regularisation is a direct consequence of the thermoparticle states, whose spectral functions are broadened around the vacuum singularity $p^{2}=m^{2}$. For higher temperatures this broadening is enhanced, which reflects the increased interactions with the surrounding thermal medium. At some temperature the kinematic peaks are completely screened, and this indicates the point at which the scattering formalism is no longer applicable. To evaluate $F(E,\vec{P};L,\beta)$ we used an exponential thermoparticle damping factor $D_{m,\beta}(\vec{x})$, which is motivated by previous lattice studies~\cite{Lowdon:2022xcl,Bala:2023iqu,Lowdon:2024atn,Ali:2026ehk,Lowdon:2025fyb,Lowdon:2025ait}. In principle $D_{m,\beta}(\vec{x})$ is uniquely fixed by the dynamical equations~\cite{Bros:2001zs}, but in practice these factors can be directly extracted from lattice correlation function data, which in the case of QCD requires the analysis of two-point correlators of single-hadron interpolating operators~\cite{Lowdon:2022xcl,Bala:2023iqu}. \\

\noindent
The formalism derived in this work can be applied to study the finite-temperature characteristics of resonances which decay into two identical scalar particles in vacuum, for example the $\rho$ meson in QCD. This formalism can also be extended to study more complicated systems, such as resonances with non-vanishing spin, as well as those with non-identical decay products. Moreover, since $L_{\tau}$ is finite in any practical lattice simulation, the current formalism can equally be used to assess the impact of finite-$L_{\tau}$ corrections in the computation of vacuum observables. These applications and extensions will be the subject of future works.

\section*{Acknowledgements}
J.H. thanks Robert Perry and Michael Wagman for inspiring discussions and comments during an earlier stage of the work. The authors acknowledge support by the Deutsche Forschungsgemeinschaft (DFG, German Research Foundation) through the Collaborative Research Center CRC-TR 211 ``Strong-interaction matter under extreme conditions'' -- Project No. 315477589-TRR 211.

\appendix

\section{Details of the numerical procedure} 
\label{sec:Appendix}

\begin{figure}[h!]
\centering
\includegraphics[width=0.49\textwidth]{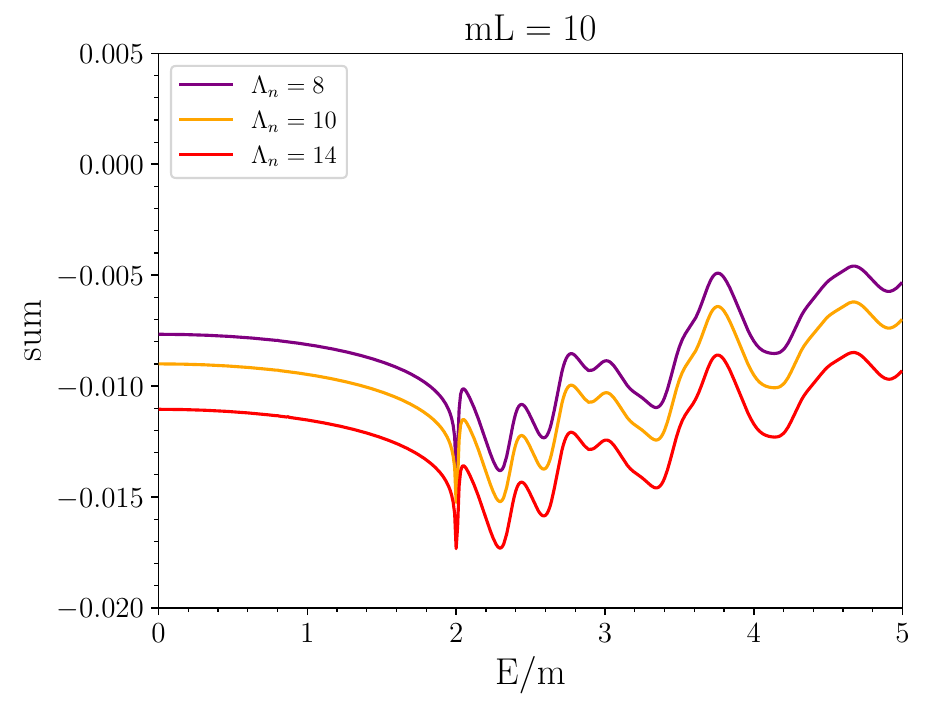}
\includegraphics[width=0.49\textwidth]{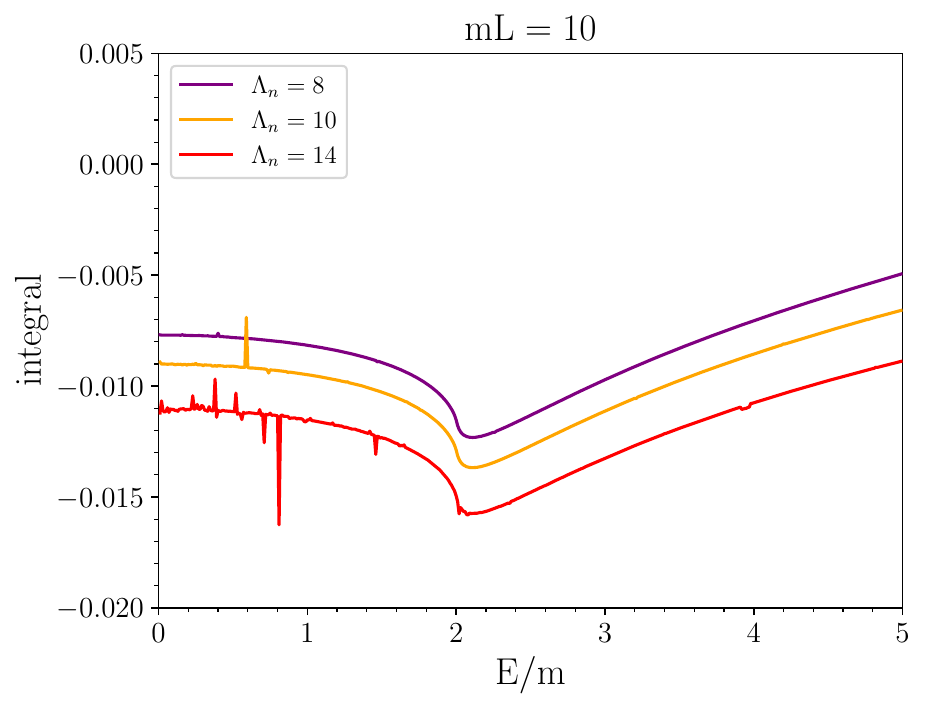}
\caption{$\vec{k}$-sum and integral components of Eq.~\eqref{eq:finite-volume-loop-principal-rest-frame} for $\gamma/m=0.1$, $mL=10$, and the spherical cutoff $\Lambda_k=\tfrac{2\pi \Lambda_n}{mL}m$ with $\Lambda_n =$ 8, 10, 14.}
\label{fig:Lüscher-L-10-sum}
\end{figure}

The first principal value integral appearing in Eq.~\eqref{eq:finite-volume-loop-principal-rest-frame} has the explicit form
\begin{align}
&\text{PV} \!\int_{m}^{\infty}\frac{dr_{0}}{2\pi}  \, \rho_{\text{TP}}(q_{0},\vec{k}) \rho_{\text{TP}}(r_{0},\vec{k})  \frac{ \coth\left(\frac{q_{0}}{2T}\right)}{|E| - q_{0}+ r_{0}} \nonumber \\
& \quad\quad\quad\quad = \lim_{\varepsilon \rightarrow 0^{+}}\left(\int_{m}^{q_{0}-|E|-\varepsilon}\frac{dr_{0}}{2\pi}+\int_{q_{0}-|E|+\varepsilon}^{\infty}\frac{dr_{0}}{2\pi}\right) \rho_{\text{TP}}(q_{0},\vec{k}) \rho_{\text{TP}}(r_{0},\vec{k})  \frac{ \coth\left(\frac{q_{0}}{2T}\right)}{|E| - q_{0}+ r_{0}},
\label{eq:PV-int}
\end{align}
where the pole contribution at $r_{0}=q_{0}-|E|$ is removed in the limit $\varepsilon \rightarrow 0^{+}$. It is not possible to take this limit after the $r_{0}$ integrals have been performed because these integrals must be evaluated numerically. Instead, the regulator $\varepsilon$ can be fixed to be sufficiently small such that finite-$\varepsilon$ corrections are negligible. In Eq.~\eqref{eq:PV-int}, as well as the other principal value integrals in Eq.~\eqref{eq:finite-volume-loop-principal-rest-frame}, the choice $\varepsilon = 10^{-6}$ met these requirements. Cutoffs $\Lambda_{r_0}=\Lambda_{q_0}=1000m$ were introduced to evaluate the infinite-domain integrals in Eq.~\eqref{eq:finite-volume-loop-principal-rest-frame}, and these values were determined to be large enough to avoid significant deviations in the numerical results. For both the $\vec{k}$-sum and integral we used the spherical cutoff $|\vec{k}|\leq \Lambda_k$, with $\Lambda_k=\tfrac{2\pi \Lambda_n}{mL}m$ and $\Lambda_n \in \mathbb{Z}^{+}$. In each numerical calculation we computed the $\vec{k}$-sum and integral for multiple values of $\Lambda_n$ in order to assess the cutoff dependence of their difference in Eq.~\eqref{eq:finite-volume-loop-principal-rest-frame}. Figure~\ref{fig:Lüscher-L-10-sum} shows the result of this cutoff separately for the sum and integral components with the parameters $\gamma/m=0.1$, $mL=10$, and $\Lambda_n =$ 8, 10, 14.  \\   

\begin{figure}[t!] 
\centering
\includegraphics[width=0.5\textwidth]{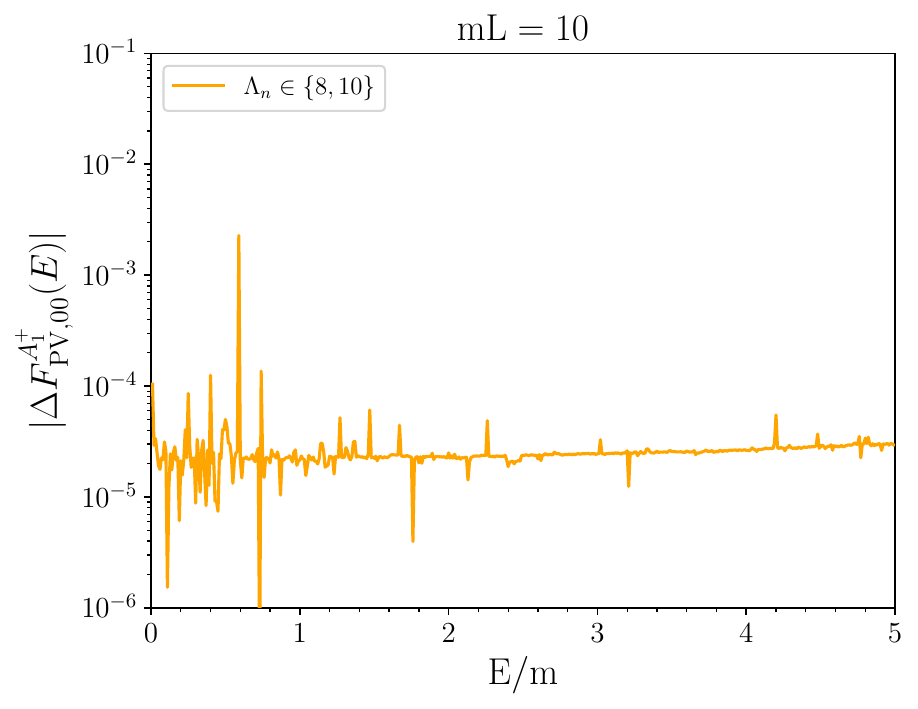}
\caption{Cutoff sensitivity $|\Delta F_{\text{PV},\,00}(E)|$ for $\gamma/m=0.1$, $mL=10$, $\Lambda_n=10$, and $\Delta \Lambda_n=2$.}
\label{fig:Lüscher-L-10-conv}
\end{figure}

\noindent
For each value of $\Lambda_n$ we chose the number of integration steps for the $q_0$ and $r_0$ integrals such that the numerical fluctuations were suppressed for $m\leq E\leq 3m$, which is the region where the most relevant two-particle energy levels are expected to lie. The sum components displayed in the left plot of Fig.~\ref{fig:Lüscher-L-10-sum} are noiseless for all energies, whilst for the integral components in the right plot one can see that fluctuations arise, particularly in the below-threshold region $E<2m$, and for larger $\Lambda_n$. These fluctuation are primarily driven by the tiny contributions from the purely thermal integrals in Eq.~\eqref{eq:finite-volume-loop-principal-rest-frame}, which are not sufficiently resolved. To improve this resolution we therefore chose a larger number of integration steps for the contributions to the $\vec{k}$-integral than for the finite-volume sum. Due to the higher computational costs for larger $\Lambda_n$ we chose $\Lambda_n$ based on the cutoff sensitivity criterion  
\begin{align}
|\Delta F_{\text{PV},\, 00}^{A_1^+}(E) | \lesssim 10^{-2}\max_{2m\leq E\leq 3m} |F_{\text{PV},\, 00}^{A_1^+}(E,\Lambda_n)|,
\label{eq:criterion}
\end{align}
where $F_{\text{PV},\, 00}^{A_1^+}(E,\Lambda_n)$ denotes the value of $F_{\text{PV},\, 00}^{A_1^+}(E)$ for a specified cutoff $\Lambda_n$, and we define 
\begin{align}
\Delta F_{\text{PV},\, 00}^{A_1^+}(E) = F_{\text{PV},\, 00}^{A_1^+}(E,\Lambda_n)-F_{\text{PV},\, 00}^{A_1^+}(E,\Lambda_n-\Delta \Lambda_n),
\end{align}
with $\Delta \Lambda_n$ a shift in the cutoff. In Fig.~\ref{fig:Lüscher-L-10-conv} the sensitivity function $|\Delta F_{\text{PV},\, 00}^{A_1^+}(E)|$ is plotted on a logarithmic scale for $\Lambda_n=10$ and $\Delta \Lambda_n=2$. The plot demonstrates that the cutoff choice $\Lambda_n=10$ meets the sensitivity criterion in Eq.~\eqref{eq:criterion} for $mL=10$. 

\newpage
  
\bibliographystyle{JHEP}

\bibliography{References} 

\end{document}